\documentclass[12pt,letterpaper]{article}
\pdfoutput=1
\usepackage{amsmath,amssymb,amsfonts}
\usepackage[T1]{fontenc}
\usepackage{palatino}
\def\author{Luca Ciambelli, Arnaud Delfante, Romain Ruzziconi, C\'eline Zwikel}
\def\title{Symmetries and Charges in Weyl--Fefferman--Graham Gauge} 
\usepackage{graphicx,nicefrac}
\usepackage[colorlinks=true, pdfstartview=FitV, linkcolor=magenta, citecolor=magenta, urlcolor=magenta]{hyperref}
\usepackage{cancel}
\usepackage{mdframed,framed}
\usepackage[usenames,dvipsnames]{xcolor}
\usepackage{mathpazo}
\usepackage{layout}
\usepackage{cite}

\newcommand{\D}{\text{d}}
\newcommand{\beq}{\begin{equation}}
\newcommand{\eeq}{\end{equation}}
\newcommand{\beqn}{\begin{eqnarray}}
\newcommand{\eeqn}{\end{eqnarray}}
\newcommand{\pa}{\partial}

\newcommand{\cO}{{\cal O}}

\usepackage{tikz-cd}
\usepackage{pict2e}

\usepackage{mathrsfs}

\numberwithin{equation}{section}

\makeatletter
\DeclareRobustCommand{\loplus}{\mathbin{\mathpalette\dog@lsemi{+}}}
\DeclareRobustCommand{\lotimes}{\mathbin{\mathpalette\dog@lsemi{\times}}}
\DeclareRobustCommand{\roplus}{\mathbin{\mathpalette\dog@rsemi{+}}}
\DeclareRobustCommand{\rotimes}{\mathbin{\mathpalette\dog@rsemi{\times}}}

\newcommand{\dog@rsemi}[2]{\dog@semi{#1}{#2}{-90,90}}
\newcommand{\dog@lsemi}[2]{\dog@semi{#1}{#2}{270,90}}
\newcommand{\dog@semi}[3]{%
  \begingroup
  \sbox\z@{$\m@th#1#2$}%
  \setlength{\unitlength}{\dimexpr\ht\z@+\dp\z@\relax}%
  \makebox[\wd\z@]{\raisebox{-\dp\z@}{%
    \begin{picture}(1,1)
    \linethickness{\variable@rule{#1}}
    \roundcap
    \put(0.5,0.5){\makebox(0,0){\raisebox{\dp\z@}{$\m@th#1#2$}}}
    \put(0.5,0.5){\arc[#3]{0.5}}
    \end{picture}%
  }}%
  \endgroup
}
\newcommand{\variable@rule}[1]{%
  \fontdimen8  
  \ifx#1\displaystyle\textfont3\else
    \ifx#1\textstyle\textfont3\else
      \ifx#1\scriptstyle\scriptfont3\else
        \scriptscriptfont3\relax
  \fi\fi\fi
}

\usepackage{mathtools}

\usepackage{tocloft}
\begin{document}

{\centering
 \vspace*{1cm}
\textbf{\LARGE{\title{}}}
\vspace{0.5cm}
\begin{center}
Luca Ciambelli$^a$, Arnaud Delfante$^b$\footnote{FRIA grantee of the Fund for Scientific Research – FNRS, Belgium.}, Romain Ruzziconi$^{c,d}$, C\'eline Zwikel$^a$\\
\vspace{0.5cm}
\textit{$^a$Perimeter Institute for Theoretical Physics,\\
31 Caroline Street North, Waterloo, Ontario, Canada N2L 2Y5}\\
\vspace{2mm}
\textit{$^b$Service de Physique de l'Univers, Champs et Gravitation,\\
Universit{\'e} de Mons -- UMONS,
20 place du Parc, 7000 Mons, Belgium} \\
\vspace{2mm}
\textit{$^c$Institute for Theoretical Physics, Technische Universit\"at Wien,\\
Wiedner Hauptstrasse 8, A-1040 Vienna, Austria} \\
\vspace{2mm}
\textit{$^d$Mathematical Institute, University of Oxford, \\ Andrew Wiles Building, Radcliffe Observatory Quarter,\\
Woodstock Road, Oxford, OX2 6GG, UK} 
\end{center}
\vspace{0.5cm}
{\small{\href{mailto:ciambelli.luca@gmail.com}{ciambelli.luca@gmail.com}, \href{mailto:Arnaud.DELFANTE@umons.ac.be}{Arnaud.DELFANTE@umons.ac.be}, \\  \href{mailto:romain.ruzziconi@tuwien.ac.at}{romain.ruzziconi@maths.ox.ac.uk}, \href{mailto:czwikel@perimeterinstitute.ca}{czwikel@perimeterinstitute.ca} }}

\vspace{1cm}
\begin{abstract}
\vspace{0.5cm}
We investigate the asymptotic symmetries of three-dimensional AdS Einstein gravity in the Weyl--Fefferman--Graham gauge, which is a  generalization of the Fefferman--Graham gauge inducing a Weyl connection as part of the boundary structure. We show that this gauge arises as a natural intermediary step of the gauge-fixing procedure in the Chern--Simons formulation. We prove that the diffeomorphism required to go to the usual Fefferman--Graham gauge can be charged, and thus its implementation has physical repercussions. We discuss the holographic renormalization and the variational principle offering a new holographic take on this gauge and its charges.
\end{abstract}}

\thispagestyle{empty}

\newpage
{
  \hypersetup{linkcolor=black}
  \tableofcontents}

\thispagestyle{empty}
\newpage
\clearpage
\pagenumbering{arabic} 


\section{Introduction} \label{sec: Introduction}

Three-dimensional Einstein gravity is a perfect playground to study gravitational physics, due to its simple features \cite{Staruszkiewicz:1963zza,Deser:1983tn,Deser:1983nh}. Indeed, albeit without propagating degrees of freedom, there are two important aspects of this theory that makes it useful. The first one is the presence of black holes \cite{Banados:1992wn,Banados:1992gq,Carlip:1995qv}, while the second is the asymptotic symmetry enhancement with respect to the isometries of the vacuum \cite{Brown:1986nw,Barnich:2001jy,Carlip:2005tz,Barnich:2007bf}. Since the precursory work of Brown and Henneaux \cite{Brown:1986nw}, it has been appreciated that AdS$_3$ with Dirichlet conditions at its conformal boundary has a centrally extended double copy Virasoro algebra of charges. This has been later interpreted as the algebra of modes of the stress tensor in the boundary conformal field theory \cite{Strominger:1997eq}, a particularly rich instance of the AdS/CFT correspondence \cite{tHooft:1993dmi,Susskind:1994vu,Maldacena:1997re,Gubser:1998bc,Witten:1998qj}. One of the advantages of three-dimensional gravity is that we can exploit its topological nature and rephrase it as a Chern--Simons (CS) theory \cite{Achucarro:1986uwr,Witten:1988hc,Banados:1994tn,Coussaert:1995zp,Henneaux:1999ib,Rooman:2000zi,Banados:2002ey,Allemandi:2002sx,Banados:2006fe,Banados:2016zim}.

The AdS/CFT correspondence is rooted in the geometric construction of the ambient space formulated by Fefferman and Graham \cite{fefferman1985conformal,Fefferman:2007rka}. These authors introduced a gauge, called Fefferman--Graham (FG) gauge, that exploits all the diffeomorphism freedom to completely lock the radial structure of the bulk metric. It is a theorem that this gauge is always attainable and, for the purposes of setting up the holographic dictionary, it has the virtue of explicitly realizing the boundary conformal multiplets as series in the holographic coordinate, see e.g. \cite{Aharony:1999ti}. It has already been noted in \cite{Ciambelli:2019bzz} that restoring some of the diffeomorphism freedom allows one to realize the Weyl connection as part of the induced boundary structure, and consequently, to restore the boundary Weyl covariance. For this reason, this partial gauge has been called the Weyl--Fefferman--Graham (WFG) gauge, which has received more attention recently \cite{Jia:2021hgy,Jia:2023gmk}. 

While not often used in the AdS/CFT correspondence, the framework of charges and the covariant phase space formalism have produced over the last decades a lot of new understanding on the concept of observables in gravity and generically in gauge theories. In particular, in the presence of boundaries of interest, it is firmly established that Noether's second theorem gives a prescription to identify the physical asymptotic symmetries 
\cite{Arnowitt:1962hi,Regge:1974zd,Benguria:1976in,Wald:1993nt,Iyer:1994ys,Wald:1999wa,Barnich:2001jy}. In gravity, the latter form the subgroup of the bulk diffeomorphisms that are compatible with the falloffs and boundary conditions and produce a non-vanishing Noether's charge, which has support on codimension-$2$ surfaces. Classifying the asymptotic symmetries is of particular interest for holography since they correspond to the symmetries of the dual theory through the holographic dictionary. Several attempts have been made to find the most general boundary conditions in three-dimensional gravity \cite{Compere:2013bya,Troessaert:2013fma,Perez:2016vqo,Grumiller:2016pqb,Grumiller:2017sjh,Campoleoni:2018ltl,Ojeda:2019xih,Grumiller:2019fmp,Adami:2020ugu,Ciambelli:2020ftk,Ciambelli:2020eba,Alessio:2020ioh,Fiorucci:2020xto,Ruzziconi:2020wrb,Geiller:2021vpg,Campoleoni:2022wmf,Adami:2023fbm,McNees:2023tus} and provide a physical interpretation from the boundary perspective. In this spirit, the corner proposal has offered an interesting angle of attack to address this problem by identifying some universal structures associated with codimension-$2$ surfaces \cite{Donnelly:2016auv,Speranza:2017gxd,Geiller:2017whh,Geiller:2017xad,Freidel:2020xyx,Freidel:2020svx,Freidel:2020ayo,Donnelly:2020xgu,Ciambelli:2021vnn,Freidel:2021cjp,Ciambelli:2021nmv,Ciambelli:2022cfr, Ciambelli:2023bmn}. Among these previous works, the results of Grumiller and Riegler \cite{Grumiller:2016pqb} stand out. In the latter, the most general solution space compatible with a well-defined variational principle has been derived. In this paper, we consider a field content which is encompassed in \cite{Grumiller:2016pqb}, but we refrain from constraining the boundary geometry. Moreover, we compute the charges in the metric formalism and show that they are finite approaching the boundary.  

One of the main results in the theory of asymptotic symmetries is that a charged diffeomorphism is a physical symmetry, mapping inequivalent physical configurations. Utilizing a charged diffeomorphism to reach a particular gauge is therefore a delicate procedure, as it restricts the physical fields of the theory. The main result of this manuscript is to show that the diffeomorphism mapping WFG to FG can be charged, and thus non-trivial. This generalizes to WFG previous enhancements of boundary conditions in FG \cite{Brown:1986nw,Detournay:2014fva,Alessio:2020ioh} and proves that even though the FG gauge is always reachable, it could constrain the physical solution space of our bulk theory. 

This opens the fascinating prospect of a more general holographic dictionary,  geometrically formulated in the WFG gauge, with extra charges and observables playing a role in the dual field theory. We explore the first steps towards this endeavour by revisiting the holographic renormalization and variational principle in WFG gauge, and identify the phase space variables with sources and VEV in the dual theory. We show that the extra physical charges coming from the partial gauge-fixing in the bulk can be modified exploiting different boundary Lagrangians and prescriptions to compute the charges, such as \cite{Compere:2008us,Papadimitriou:2005ii,Freidel:2020xyx, Compere:2020lrt, McNees:2023tus}.
In particular, we focus on three different symplectic structures. In the first one, there are no extra charges with respect to the FG gauge, and thus Weyl-covariantization can be attained for free. In the second, the Weyl connection appears as a source in the boundary dual theory, but the renormalization procedure is not covariant. Lastly, we propose a symplectic structure that gives rise to a boundary Weyl-invariant extra source (on top of the boundary metric), with a covariant renormalization procedure. While all of them are mathematically sound, they lead to different physical
interpretations and consequences for the boundary dual theory.

The paper is organized as follows. We start in section \ref{sec: GFG} revisiting the gauge fixing procedure in the CS formulation. We explain how WFG naturally arises as an intermediate step in the gauge-fixing toward the FG gauge. We then comment on why the WFG gauge is useful, reviewing the appearance of Weyl symmetry. Section \ref{sec: Higher order} contains the novelties of this paper. Specifically, we compute the asymptotic symemtries and charges, showing explicitly that the diffeomorphism bringing WFG to FG is charged. Section \ref{sec:HolographicWI} discusses some holographic aspects of the WFG gauge. The variational principle is presented and the holographic Ward identities are derived. We then conclude offering possible outlooks in section \ref{sec: Conclusions}. Two appendixes are offered, where we report conventions and some technical details used in the main body of the manuscript.

\section{Gauge fixing procedure} \label{sec: GFG}

In this section, we introduce the Weyl--Fefferman--Graham (WFG) gauge \cite{Ciambelli:2019bzz}. In subsection \ref{subsec: FG to WFG}, we begin by studying a natural way to relax the Fefferman--Graham (FG) gauge using the Chern--Simons (CS) formulation, realizing the radially leading part of the WFG gauge. This will turn out to be sufficient for the asymptotic analysis we are interested in.   We then review in subsection \ref{subsec: Why WFG} the results of \cite{Ciambelli:2019bzz,Jia:2021hgy,Jia:2023gmk}, where the WFG gauge was introduced to  restore the Weyl covariance of the boundary theory. 

\subsection{From FG to WFG} \label{subsec: FG to WFG}
In this section, we relax the gauge conditions leading to the Fefferman--Graham (FG) gauge. We show that there is a natural intermediate gauge obtained in this way, which is the Weyl--Fefferman--Graham (WFG) gauge where only the  Weyl connection, appearing at the leading radial order of the bulk metric expansion, is turned on. This relaxation was initially studied in \cite{Grumiller:2016pqb, Ciambelli:2019bzz} and further analyzed in \cite{Jia:2021hgy,Jia:2023gmk}.

The Fefferman--Graham gauge of $3$-dimensional anti-de Sitter space consists in choosing bulk coordinates $x^\mu = (\rho,x^i)$, where $\rho \geq 0$ is a radial coordinate and $x^i = (t,\theta)$ the boundary coordinates such that the boundary is located at $\rho = 0$, as well as the line element \cite{Starobinsky:1982mr,fefferman1985conformal,Fefferman:2007rka,Skenderis:2002wp,Papadimitriou:2010as}\footnote{We follow conventions similar to \cite{Ruzziconi:2019pzd,Ciambelli:2020ftk,Ciambelli:2020eba}.}
\begin{equation} \label{FG ansatz of the bulk metric}
	\text{d}s^2_{\text{FG}} = g_{\mu\nu} \text{d}x^\mu \text{d}x^\nu = \frac{\text{d}\rho^2}{\rho^2}  + h_{ij}(\rho,x) \text{d}x^i \text{d}x^j \, ,
\end{equation}
where we have fixed the AdS radius, $\ell=1$. The rewriting of this line element in the Chern--Simons formulation leads to a natural relaxation of this gauge. We refer to appendix \ref{app sec: Conventions CS} for a review of the CS formalism. In this context, the $\mathfrak{sl}(2,\mathbb{R})$ connection decomposes as
\begin{equation}
    A = A_\mu \text{d}x^\mu = A_\rho \text{d}\rho + A_i \text{d}x^i \, ,
\end{equation}
and similarly for $\widetilde{A}$. The associated CS action is invariant under SL(2,$\mathbb{R}$) gauge transformations
\begin{equation} \label{SL2R gauge transformation}
    A \to U^{-1} A U + U^{-1} \text{d} U \, ,
\end{equation}
whose infinitesimal version is
\begin{equation} \label{gauge symmetry of CS field}
    \delta_\lambda A = I_{V_\lambda} \delta A = \text{d}\lambda + [A,\lambda] \, ,
\end{equation}
where $\lambda \in \mathfrak{sl}(2,\mathbb{R})$ and $U = \text{exp}(\lambda)$ is its related group element. Depending on the holographic description we want to study, we can exploit this gauge invariance to select a specific gauge for the CS fields $A$. An arbitrary radial expansion is treated in subsection \ref{subsec: Why WFG} and in section \ref{sec: Higher order}. In this subsection we consider a gauge leading to a finite $\rho$-expansion, as the Fefferman--Graham one. Indeed, if we impose that
\begin{equation}
    A_\rho = - \frac{1}{\rho} \, L_0 \, , \qquad \widetilde{A}_\rho = \frac{1}{\rho} \, L_0 \, ,
\end{equation}
we get the following FG gauge-fixing condition
\begin{equation}
    g_{\rho\rho} = - \text{tr} \left( A_\rho - \widetilde{A}_\rho \right)^2 = \frac{1}{\rho^2} \, ,
\end{equation}
where we employed the following $\mathfrak{sl}(2,\mathbb{R})$ basis\footnote{A possible matrix realization of this basis is
\begin{equation}
    L_1 = - \frac{1}{\sqrt{2}} \begin{pmatrix} 0 & 0 \\ 1 & 0 \end{pmatrix} \, , \qquad L_{-1} = - \frac{1}{\sqrt{2}} \begin{pmatrix} 0 & 1 \\ 0 & 0 \end{pmatrix} \, , \qquad L_0 = \frac{1}{2} \begin{pmatrix} 1 & 0 \\ 0 & -1 \end{pmatrix} \, .
\end{equation}}
\begin{equation} \label{sl(2,R)}
    [L_1,L_{-1}] = - L_0 \, , \qquad [L_1,L_0] = L_1 \, , \qquad [L_{-1},L_0] = - L_{-1} \, .
\end{equation}

In three spacetime dimensions, the on-shell FG line element has a truncated radial expansion. This motivates the radial gauge choice for the CS connections
\begin{equation} \label{first CS gauge choice}
    A = b^{-1} \, a \, b + b^{-1} \, \text{d} b \, , \qquad \widetilde{A} = \widetilde{b}^{-1} \, \widetilde{a} \, \widetilde{b} + \widetilde{b}^{-1} \, \text{d} \widetilde{b} \, .
\end{equation}
The one-forms $a = a_i \text{d}x^i$ and $\widetilde{a} = \widetilde{a}_i \text{d}x^i$ encode then all the information contained in the metric. The group elements $b$ and $\widetilde{b}$ are fixed to satisfy
\begin{equation}
    A_\rho = b^{-1} \, \partial_\rho b \, , \qquad \widetilde{A}_\rho = \widetilde{b}^{-1} \, \partial_\rho \widetilde{b} \, ,
\end{equation}
leading to
\begin{equation} \label{WFG CS radial group element}
    b = \text{exp}\left(-\log \rho \, L_0\right) \, , \qquad \widetilde{b} = b^{-1} \, .
\end{equation}
We note that this gauge choice is always possible (at least locally): one can always find such a group element $U$ that gives the identity at the boundary, see \cite{Campoleoni:2010zq}. Using one of the equations of motion, that is,
\begin{equation}
    F_{\rho i} = \partial_\rho A_i + A_\rho \, A_i - \left( \rho \leftrightarrow i \right) = 0 \, ,
\end{equation}
the $\mathfrak{sl}(2,\mathbb{R})$-valued one-forms $a$ and $\widetilde{a}$ depend only on the boundary coordinates. This fact constitutes a considerable advantage in the calculation of surface charges, compared to the metric formalism. Indeed, we will not need to recur to holographic renormalization, because the radial dependence is completely captured by the gauge transformation and factors out. We will comment on this last fact below eq. \eqref{total corner charge}. The remaining equations of motion simply tell us that these connections are flat,
\begin{equation}
    \text{d}a + a \wedge a = 0 \, .
\end{equation}

If we expand them in the above basis \eqref{sl(2,R)} as
\begin{equation} \label{a expanded in sl2}
    a_i = a_i^1 \, L_1 + a_i^0 \, L_0 + a_i^{-1} \, L_{-1} \, ,
\end{equation}
and similarly for $\widetilde{a}$, one can deduce that
\begin{equation} \label{grhoi CS before FG gauge}
    g_{\rho i} = - \frac{1}{2\rho} \left(a_i^0 - \widetilde{a}_i^0 \right) .
\end{equation}
Therefore, without further restriction, this does not verify the FG gauge-fixing condition $g_{\rho i} = 0$. This then leads us to a natural relaxation of the FG gauge \eqref{FG ansatz of the bulk metric}, where the bulk metric takes the form
\begin{equation} \label{WFG ansatz of the bulk metric}
    g_{\mu\nu} \text{d}x^\mu \text{d}x^\nu = \Big( \frac{\text{d}\rho}{\rho} - k_i(\rho,x) \text{d}x^i \Big)^2 + h_{ij}(\rho,x) \text{d}x^i \text{d}x^j \, ,
\end{equation}
such that
\begin{equation} \label{WFG CS radial expansion of ki}
    k_i(\rho,x) = k_i^{(0)}(x) = \frac{1}{2} \left( a_i^0 - \widetilde{a}_i^0 \right),
\end{equation}
and
\begin{subequations} \label{WFG CS radial expansion of hij}
    \begin{align}
        &h_{ij}(\rho,x) = \frac{1}{\rho^2} h_{ij}^{(0)}(x) + h_{ij}^{(2)}(x) + \rho^2 h_{ij}^{(4)}(x) \, ,\\
        &h_{ij}^{(0)} = - \frac{1}{4} \, a_i^1 \, \widetilde{a}_j^{-1} \, , \qquad h_{ij}^{(2)} = \frac{1}{4} \left( a_i^1 \, a_j^{-1} + \widetilde{a}_i^{-1} \, \widetilde{a}_j^{1} \right) , \qquad h_{ij}^{(4)} = - \frac{1}{4} \, a_i^{-1} \, \widetilde{a}_j^{1} \, .
    \end{align}
\end{subequations}
It is worth noting that the gauge choice \eqref{WFG ansatz of the bulk metric} can be imposed off-shell as the starting point, as we will do in section \ref{sec: Higher order}. We will also show that the line element \eqref{WFG ansatz of the bulk metric}  can be obtained from \eqref{FG ansatz of the bulk metric} acting with the diffeomorphism \eqref{Weyl in WFG}.

This gauge was first introduced in \cite{Banados:1994tn} in the Chern--Simons formulation, and systematically analysed in \cite{Grumiller:2016pqb}. It was studied geometrically in \cite{Ciambelli:2019bzz} in the metric formalism for any dimension. In the latter, it was named Weyl--Fefferman--Graham (WFG) since it was shown that it induces a Weyl connection as part of the boundary geometry. We will review this in subsection \ref{subsec: Why WFG}. We note that \eqref{WFG CS radial expansion of ki} incorporates only the zero radial order of the new structure $k_i(\rho,x)$. In reference \cite{Grumiller:2016pqb}, it has been proposed to choose another group element instead of \eqref{WFG CS radial group element}, i.e.,
\begin{equation}
    b(\rho) = \text{exp}(L_{-1}) \text{exp}(-\log(\rho) L_0) \, ,
\end{equation}
leading to
\begin{equation}
	k_i(\rho,x) = \frac{1}{\rho} k_i^{(-1)}(x) + k_i^{(0)}(x) + \rho k_i^{(1)}(x) \, ,
\end{equation}
while in \cite{Ciambelli:2019bzz} the following infinite expansion was considered:
\begin{equation}
    k_{i}(\rho,x) = \sum_{n \ge 0} \rho^{2n} \, k_{i}^{(2n)}(x) \, .
\end{equation}

In the following, we will focus on the latter. In particular, in subsection \ref{subsec: Why WFG} we will geometrically motivate the introduction of the WFG gauge via the second order formalism. On the other hand, in section \ref{sec: Higher order} we will come back to the first order formalism and determine whether the new fields introduced have associated non-vanishing charges or not. This is an important and delicate step, because one has to determine whether restricting the radial expansion \eqref{first CS gauge choice} to be finite is a pure gauge choice, or instead it is  a physical restriction on the asymptotic phase space.   Moreover, from a CS viewpoint, it is always possible to go back to the FG metric \eqref{FG ansatz of the bulk metric} by performing a residual gauge transformation which yields the following gauge fixing \cite{Banados:2004nr}: 
\begin{equation} \label{diffeomorphism required to go the usual FG gauge}
    a_i^0 = \widetilde{a}_i^0 \qquad \Rightarrow \qquad g_{\rho i} = - \frac{1}{\rho} k_i = 0 \, .
\end{equation}
This has been shown in \cite{Banados:2002ey} and is related to the FG theorem \cite{Fefferman:2007rka}. However, as just stated, the radial and FG gauge fixing transformations can be physical, which is determined via the study of surface charges in this theory.

\subsection{Why WFG} \label{subsec: Why WFG}

In this subsection, we briefly review the WFG geometry \cite{Ciambelli:2019bzz,Jia:2021hgy,Jia:2023gmk}. Even though the FG gauge \eqref{FG ansatz of the bulk metric} is convenient in the AdS/CFT correspondence, it breaks the explicit Weyl covariance of the boundary by inducing the Levi-Civita connection from the bulk. The WFG gauge was precisely introduced to restore the Weyl covariance of the boundary.

In the FG gauge, a radial rescaling must be accompanied by a diffeomorphism in the transverse space, so to preserve the gauge. Such a transformation is called a Penrose--Brown--Henneaux (PBH) transformation \cite{Penrose:1985bww,Brown:1986nw,Imbimbo:1999bj,Rooman:2000zi,Rooman:2000ei,Bautier:2000mz,Ciambelli:2019bzz}, and has the form
\begin{equation} \label{PBH transformation}
	\rho \to \rho ' = \frac{\rho}{\mathfrak{B}(x)} \, , \qquad x^i \to {x'}^i = x^i + \xi^i(\rho,x) \, .
\end{equation}
Therefore, if we want to induce a Weyl transformation on the boundary in the FG gauge we have to simultaneously introduce the diffeomorphism $\xi^i(\rho,x)$, which vanishes at the boundary $\rho = 0$. This transformation impacts the subleading terms in the asymptotic radial FG expansion so that they do not transform Weyl-covariantly under \eqref{PBH transformation}.

This issue can be overcome by relaxing the FG ansatz to the WFG gauge \cite{Ciambelli:2019bzz}
\begin{equation} \label{WFG ansatz of the bulk metric 2}
	\text{d}s^2_{\text{WFG}} = g_{\mu\nu} \text{d}x^\mu \text{d}x^\nu = \Big( \frac{\text{d}\rho}{\rho} - k_i(\rho,x) \text{d}x^i \Big)^2 + h_{ij}(\rho,x) \text{d}x^i \text{d}x^j \, .
\end{equation}
In the last equation, we introduced the quantities $k_i$ and $h_{ij}$, which can be asymptotically radially expanded as
\begin{equation} \label{WFG expansion}
	h_{ij}(\rho,x) = \frac{1}{\rho^2} \sum_{n \ge 0} \rho^{2n} \, h_{ij}^{(2n)}(x) \, , \qquad k_{i}(\rho,x) = \sum_{n \ge 0} \rho^{2n} \, k_{i}^{(2n)}(x) \, ,
\end{equation}
where $h_{ij}^{(0)}$ is the boundary metric and $k_i^{(0)}$ is a boundary Weyl connection. In the following, we will review why this is the right interpretation. It is worth noting that \eqref{WFG ansatz of the bulk metric 2} is the metric \eqref{WFG ansatz of the bulk metric} we obtained by a natural gauge relaxation in the CS formulation, but with an arbitrary radial dependence of $k_i$. As we will explain later, this arbitrary radial dependence can be fixed to recover exactly \eqref{WFG ansatz of the bulk metric} by using pure-gauge residual diffeomorphisms. 

We start observing that the form of \eqref{WFG ansatz of the bulk metric 2} is now preserved under the following radial diffeomorphisms inducing boundary Weyl transformations;
\begin{equation}
	\rho \to \rho ' = \frac{\rho}{\mathfrak{B}(x)} \, , \qquad x^i \to {x'}^i = x^i \, , \label{Weyl in WFG}
\end{equation}
so that
\begin{subequations}
    \begin{align}
	&k_i(\rho,x) \to {k'}_i(\rho ', x') = k_i(\mathfrak{B}(x)\rho ',x) - \partial_i \text{ln} \mathfrak{B}(x) \, ,\\
        &h_{ij}(\rho,x) \to {h '}_{ij}(\rho ', x) = h_{ij}(\mathfrak{B}(x) \rho ',x) \, .
    \end{align}
\end{subequations}
This then solves the above highlighted problem, because the subleading terms in the radial expansion are now covariant under Weyl transformations. Thanks to the last equations, we can also see that
\begin{equation}
	k_i^{(2n)}(x) \to k_i^{(2n)}(x) \, \mathfrak{B}(x)^{2n} - \delta_{n,0} \, \partial_i \text{ln} \mathfrak{B}(x) \, , \qquad h_{ij}^{(2n)}(x) \to h_{ij}^{(2n)}(x) \, \mathfrak{B}(x)^{2n-2} \, .
\end{equation}
Except for the leading term in $k_i$, all terms in the $\rho$ expansion transform as Weyl tensors, with a definite Weyl weight given by the power of $\mathfrak{B}(x)$. The leading term in $k_i$, on the other hand, transforms inhomogeneously, i.e., as a Weyl connection,
\begin{equation} \label{Weyl transformation of the Weyl connection}
	k_i^{(0)} \to k_i^{(0)} - \pa_i \text{ln} \mathfrak{B} \, .
\end{equation}
One also notices that the leading order of $h_{ij}$, the boundary metric, varies as
\begin{equation}
	h_{ij}^{(0)} \to \mathfrak{B}^{-2} h_{ij}^{(0)} \, .
\end{equation}

In the following, we define a Weyl-covariant derivative in the boundary that parallel transports the boundary metric $h^{(0)}_{ij}$.
The idea is to choose the following dual form basis\footnote{It should not be confused with the dreibein used in the first order formulation as in, e.g., eq. \eqref{WFG ansatz of the bulk dreibein}.} of the metric \eqref{WFG ansatz of the bulk metric 2}
\begin{equation}
	E^\rho = \frac{\text{d}\rho}{\rho} - k_i(\rho,x) \text{d}x^i \, , \qquad E^i = \text{d}x^i \, ,
\end{equation}
and its corresponding vector basis
\begin{equation} \label{Vector basis of the tangent space}
	E_\rho = \rho \partial_\rho \equiv D_\rho \, , \qquad E_i = \partial_i + \rho k_i(\rho,x) \partial_\rho \equiv D_i \, .
\end{equation}
The basis $\{D_\rho,D_i\}$ spans the tangent space at any point $(\rho,x^i)$ of the bulk manifold $\mathcal{M}$, and the spatial vectors $\{D_i\}$ span a $2$-dimensional distribution $\mathcal{D}$ on $\mathcal{M}$. Their Lie brackets are given by
\begin{equation}
	[D_\rho,D_i] = D_\rho k_i \, D_\rho \, , \qquad [D_i,D_j] = f_{ij} \, D_\rho \, ,
\end{equation}
where
\begin{equation} \label{new Weyl curvature}
    f_{ij} \equiv D_i k_j - D_j k_i
\end{equation}
is the curvature associated to $k_i$. Using the standard definition, the coefficients of the bulk Levi-Civita (LC) connection $\nabla$ in the frame $\{D_\rho,D_i\}$ are
\begin{equation}
	\nabla_{D_i} D_j = \Gamma^{k}_{ij} D_k + \Gamma^\rho_{ij} D_\rho \, .
\end{equation}
Taking into account the radial expansions \eqref{WFG expansion} to compute $\Gamma^{k}_{ij}$, one obtains 
\begin{equation} \label{Zero order coefficient connection}
	({\Gamma^{(0)}})^k_{ij} = \frac{1}{2} h^{kl}_{(0)} \left( (\partial_i-2 k^{(0)}_i)h_{jl}^{(0)} + (\partial_j-2 k^{(0)}_j) h_{il}^{(0)} - (\partial_l-2 k^{(0)}_l) h_{ij}^{(0)} \right),
\end{equation}
at leading order. We emphasize that the coefficients $\Gamma^{k}_{ij}$ define the induced connection coefficients on $\mathcal{D}$, see e.g. \cite{Lecanda2018}, while their radial zero order \eqref{Zero order coefficient connection} provides the coefficients of a torsion-free connection with Weyl metricity \cite{folland1970weyl,hall1992weyl, Ciambelli:2019bzz}. 

Therefore, the WFG gauge has the novelty of being equipped with a Weyl geometry and a metric $h_{ij}^{(0)}$ at the boundary. The induced connection $\nabla^{(0)}$ acts as follows,
\begin{equation} \label{Weyl boundary connection}
	\nabla_i^{(0)} h_{jk}^{(0)} = 2 k_i^{(0)} h_{jk}^{(0)} \, .
\end{equation}
For a generic Weyl-weight $\omega_T$ tensor $T$ of arbitrary type, we can construct the Weyl covariant connection as
\begin{equation}
	\hat{\nabla}_i^{(0)} T \equiv \nabla_i^{(0)} T + \omega_T k_i^{(0)} T \, . \label{WeylcovD}
\end{equation}
So the connection $\hat{\nabla}^{(0)}$ is metric and $\hat{\nabla}_i^{(0)} T$ is Weyl covariant. Moreover, thanks to this, all geometric quantities built with this connection are Weyl covariant. 

We end this subsection with an important remark. If the Weyl curvature $f_{ij}$ is zero, that is, $[D_i,D_j] = 0$, the distribution $\mathcal{D}$ is integrable according to the Frobenius theorem. In FG gauge $\mathcal{D}$ becomes $\{ \partial_i \}$ and foliates $\mathcal{M}$ with $\rho$-constant surfaces, which is not necessarily true for WFG.

\section{Asymptotic symmetries} \label{sec: Higher order}

In this section, we come back to the Chern--Simons formulation of the WFG gauge \eqref{WFG ansatz of the bulk metric 2}. More specifically, in subsection \ref{subsec:WFG solution space first order} we determine the bulk asymptotic solution space in a conformal parametrization of the boundary, while we compute the asymptotic symmetries and the associated surface charges in subsection \ref{subsec:WFG charges first order}. We show that the transformation \eqref{diffeomorphism required to go the usual FG gauge}, which further gauge fixes WFG down to FG, is charged, whereas the radial gauge fixing \eqref{first CS gauge choice} is a pure gauge diffeomorphism.

\subsection{Solution space} \label{subsec:WFG solution space first order}

We derive the asymptotic solution space in the WFG gauge \eqref{WFG ansatz of the bulk metric 2} in the first order formalism. 
Using $g_{\mu\nu}={e_\mu}^B \eta_{BC} \, {e_\nu}^C$, with  ($B = +1, -1, 0$), we consider the following dreibein
\begin{equation} \label{WFG ansatz of the bulk dreibein}
	e^0=e_{\mu}{}^0\text{d}x^\mu = - \frac{\text{d}\rho}{\rho} +  k_i(\rho,x) \text{d}x^i\, , \qquad e^{\pm 1} =e_{\mu}{}^{\pm 1}\text{d}x^\mu = \frac{1}{\rho} {e_i}^{\pm 1}(\rho,x) \text{d}x^i \, ,
\end{equation}
where we employed the basis \eqref{sl(2,R)} and the Minkowski metric
\begin{equation}
	\eta_{BC} = 2 \, \text{tr} \left( L_B \, L_C \right) = \begin{pmatrix} 0 & 1 & 0 \\ 1 & 0 & 0 \\ 0 & 0 & 1 \end{pmatrix} .
\end{equation}
Using \eqref{WFG expansion}, we have the following radial expansion
\begin{equation} \label{WFG CS generic expansion}
	k_{i}(\rho,x) = \sum_{n \ge 0} \rho^{2n} \, k_{i}^{(2n)}(x) \, , \qquad {e_i}^{\pm 1}(\rho,x) = \sum_{n \ge 0} \rho^{2n} ({e_i}^{\pm 1})^{(2n)}(x) \, .
\end{equation}
One can determine the associated spin connection and the CS forms using the formulas collected in appendix \ref{app sec: Conventions CS}, in particular \eqref{Cartan equation} and \eqref{A-E}. It is then possible to solve the flatness conditions reported in \eqref{eom CS} order by order in the radial expansion. Since the boundary is two-dimensional, we can always express its metric in a conformally-flat parametrization
\begin{equation} \label{conformal gauge condition bdy metric}
	h_{ij}^{(0)}(x) = \text{e}^{2 \phi(x)} \eta_{ij} \, ,
\end{equation}
where $\phi$ is an arbitrary conformal factor and $\eta_{ij}$ is the two-dimensional Minkowski metric. We introduce the light-cone coordinates
\begin{equation} \label{light-cone coordinates}
	x^\pm =  \pm t+ \theta \, , 
\end{equation}
for which $\partial_t=\partial_+-\partial_-$ and $\partial_\theta=\partial_++\partial_-$.
The boundary metric takes the form
\begin{equation}
	\text{d}s^2_{\text{bdy}} = h_{ij}^{(0)} \text{d}x^i \text{d}x^j = \text{e}^{2 \phi(x^+,x^-)} \text{d}x^+ \text{d}x^- \, .
\end{equation}
The corresponding bulk metric is then
\begin{equation} \label{WFG ansatz of the bulk metric 3}
    \text{d}s^2_{\text{bulk}} = g_{\mu\nu} \text{d}x^\mu \text{d}x^\nu = \Big( \frac{\text{d}\rho}{\rho} - k_i(\rho, x) \text{d}x^i \Big)^2 + h_{ij}(\rho,x) \text{d}x^i \text{d}x^j \, ,
\end{equation}
where 
\begin{equation}
    h_{ij}(\rho,x) = \frac{1}{\rho^2} h_{ij}^{(0)}(x) + h_{ij}^{(2)}(x) + \rho^2 h_{ij}^{(4)}(x) + \cO(\rho^4)
\end{equation}
such that
\begin{subequations}
\begin{align}
	h_{\pm \pm}^{(0)} &= 0 \, ,\\
	h_{\pm \pm}^{(2)} &= \ell_\pm - (K_\pm^{(0)})^2 - \partial_\pm K_\pm^{(0)} ,\\
    \begin{split}
	h_{\pm \pm}^{(4)} &= - \text{e}^{-2\phi} \, \partial_\pm K_\mp^{(0)} h_{\pm \pm}^{(2)} - k_\pm^{(2)} \left( \partial_\pm \phi + 2 K_\pm^{(0)} \right) - \frac{1}{2} \partial_\pm k_\pm^{(2)} \, , 
    \end{split}
\end{align}
\end{subequations}
and
\begin{subequations}
\begin{align}
	h_{+-}^{(0)} &= \frac{1}{2} \text{e}^{2 \phi} \, ,\\
	h_{+-}^{(2)} &= - \frac{1}{2} \Big( \partial_- K_+^{(0)} + \partial_+ K_-^{(0)} \Big) \, ,\\
	\begin{split}
	h_{+-}^{(4)} &= \frac{1}{4} \text{e}^{-2\phi} \Big[  2\partial_+ K_-^{(0)} \partial_- K_+^{(0)} - \text{e}^{2\phi} \Big( \partial_- k_+^{(2)} + \partial_+ k_-^{(2)} + 2 k_+^{(0)} k_-^{(2)} + 2 k_-^{(0)} k_+^{(2)} \Big) + 2 h_{++}^{(2)} h_{--}^{(2)} \Big] \, .
	\end{split}
\end{align}
\end{subequations}
Here we defined 
\begin{equation} \label{K_pm^0}
    K_\pm^{(0)} = k_\pm^{(0)} - \partial_\pm \phi \, ,
\end{equation}
which is the Weyl gauge connection shifted by a pure gauge factor, see \eqref{Weyl transformation of the Weyl connection}. Moreover, the equation of motions impose \begin{equation}\partial_\pm \ell_\mp=0 \, .\end{equation}
Note that the higher orders $h_{ij}^{(2n)}$ depend on $k_i^{(2n)}$. 
Just like in \cite{Alessio:2020ioh} and as opposed to \cite{Troessaert:2013fma}, we also notice that we do not perform a chiral split of the solution space, because we consider here arbitrary boundary functions $\phi$ and $k_i$. We can derive the following flat CS connections
\begin{subequations} \label{boundary conditions on A - CS WFG zero order conformal gauge}
\begin{align}
    A_\rho &= - \frac{1}{\rho} L_0 + 2 \sqrt{2} \, \rho^2 \text{e}^{-\phi} \left( k_-^{(2)} L_1 - k_+^{(2)} L_{-1} \right)+\cO(\rho^3) \, ,\\
    A_+ &= \frac{\sqrt{2}}{\rho} \text{e}^\phi L_1 + \left( 2 \, k_+^{(0)} - \partial_+ \phi \right) L_0 + \sqrt{2} \, \rho \, \text{e}^{-\phi} h_{++}^{(2)} L_{-1} + 2 \, \rho^2 k_+^{(2)} L_0 + \cO(\rho^3) \, ,\\
    A_- &= \partial_- \phi L_0 - \sqrt{2} \, \rho \, \text{e}^{-\phi} \partial_- K_+^{(0)} L_{-1}+ \cO(\rho^3) \, .
\end{align}
\end{subequations}
A straightforward observation is that, unlike the FG gauge in three dimensions, the WFG gauge has an infinite radial expansion. If we turn on only the leading order of the Weyl connection, the expansion reduces to a finite sum. As customary in the literature, we focus on the first $\mathfrak{sl}(2,\mathbb{R})$ copy in the next steps. We collect the main results on the second copy in appendix \ref{app sec: Second copy sl(2,R)}. 

To summarize the solution space is given by one free function parametrizing the boundary metric $\phi(x^\pm)$, the Weyl connection $k_i(\rho,x^\pm)$ and two chiral functions $\ell^{\pm}(x^\pm)$ whose zero mode encode a combination of the mass and the angular momentum. 

\subsection{Charges} \label{subsec:WFG charges first order}
We now turn to the identification of the residual symmetries of the solution space and their associated surface charges. 

\paragraph{Residual symmetries}

We have to determine the gauge transformations \eqref{gauge symmetry of CS field} preserving the form of the boundary values of $A$ in \eqref{boundary conditions on A - CS WFG zero order conformal gauge}. We assume the following basis expansion for the gauge parameters
\begin{equation} \label{expansion_parameter}
\lambda(\rho,x^+,x^-) = \epsilon^B(\rho,x^+,x^-) L_B \, ,
\end{equation}
where the basis is introduced in \eqref{sl(2,R)}.
Then the components are given by
\begin{subequations} \label{ChernSimons - A gauge parameters}
    \begin{align}
        \begin{split}
        \epsilon^1 &= \frac{\sqrt{2}}{\rho} \text{e}^{\phi} Y^+ +\cO(\rho^3) \, ,
        \end{split}\\
        \begin{split}
        \epsilon^{-1} &= \frac{\rho}{\sqrt{2}} \text{e}^{-\phi} \Big[ \partial_+^2 Y^+ - 2 H_+^{(0)} + 2 \Big( \ell_+ Y^+ -(K_+^{(0)})^2 Y^+ + K_+^{(0)} \partial_+ Y^+ \Big) \Big]  +\cO(\rho^3) \, ,
        \end{split}\\
        \epsilon^0 &= \sigma - \partial_+ Y^+ + 2 Y^+ K_+^{(0)} \, ,
    \end{align}
\end{subequations}
where $\sigma$ and $H_\pm^{(0)}$ are arbitrary functions of the boundary coordinates while $Y^\pm = Y^\pm(x^\pm)$.
In analogy with \eqref{K_pm^0}, we introduce $h_\pm^{(0)}$ defined via
\begin{equation}
    H_\pm^{(0)} = h_\pm^{(0)} - \partial_\pm \sigma \, . \label{field dep redef}
\end{equation}
Then, the asymptotic Killing vectors are given by\footnote{These are obtained using $\xi^\mu =\tfrac{1}{2}  {e_B}^\mu ( \lambda^B - \widetilde{\lambda}^B )$, see \cite{Witten:1988hc}.}
\begin{equation} \label{AKV CS WFG}
	\xi^\rho = \rho \, \omega + \mathcal{O}(\rho^3) \, , \qquad \xi^\pm = Y^\pm + \rho^2 \zeta^\pm + \mathcal{O}(\rho^4) \, ,
\end{equation}
where we renamed
\begin{subequations}
    \begin{align}
	\omega(x^+,x^-) &= - \sigma + \frac{1}{2} \partial_i Y^i+ Y^i \partial_i \phi\, ,\\
	\zeta^\pm(x^+,x^-) &= \text{e}^{-2\phi} \left(K^{(0)}_\mp\pa_\mp Y^\mp-H^{(0)}_{\mp}-\frac1{2}\pa_\mp \pa_i Y^i+Y^i\pa_i K^{(0)}_\mp\right),
	\end{align}
\end{subequations}
such that $\xi^\rho$ and $\xi^\pm$ have parameters $\omega$ and $H^{(0)}_\pm$ that do not mix.

In particular, we see that the Weyl parameter $\omega$ is given by
\begin{equation}\label{o}
	\omega - k_i^{(0)} Y^i= - \frac{1}{2} \left( \epsilon^0 - \widetilde{\epsilon}^0 \right) ,
\end{equation}
which shows a link between radial rescalings and $\mathfrak{sl}(2,\mathbb{R})$ gauge transformations along the Cartan direction $L_0$. Indeed, the bulk radial rescaling \eqref{PBH transformation} acts on the boundary metric as \cite{Imbimbo:1999bj}
\begin{equation}
    h_{ij}^{(0)} \to \text{e}^{2 \omega} h_{ij}^{(0)} \, ,
\end{equation}
that is, it induces a Weyl transformation with parameter $\omega(x)$. We note that our expression \eqref{o} is Weyl covariant, since the standard relation would rather be $\omega = - \tfrac{1}{2} ( \epsilon^0 - \widetilde{\epsilon}^0)$, see \cite{Banados:2004nr,Rooman:2000zi,Li:2015osa}.
Once the residual gauge transformations are identified, we can compute their commutation relations. Since these generators are field dependent, one has to use a modified commutator\footnote{It is analogous to its metric counterpart: the residual diffeomorphism algebra is closed under the modified Lie bracket of two asymptotic Killing vectors \cite{Schwimmer:2008yh, Barnich:2010eb}
\begin{equation} \label{modified-Lie}
    [\xi_1,\xi_2]_M \equiv [\xi_1,\xi_2]-\delta_{\xi_1}\xi_2+\delta_{\xi_2}\xi_1 \, ,
\end{equation}
where $[\xi_1,\xi_2]$ is the standard Lie bracket. The modified bracket takes into account the dependence of the vectors on fields that transform themselves under the symmetry.}
\begin{equation} \label{ChernSimons - Modified Commutator of Gauge Parameters}
[\lambda_1,\lambda_2]_M \equiv [\lambda_1,\lambda_2] - \delta_{\lambda_1} \lambda_2 + \delta_{\lambda_2} \lambda_1 \, ,
\end{equation}
where, e.g., $\delta_{\lambda_1} \lambda_2$ denotes the variation of $\lambda_2$ under $\lambda_1$. A similar relation applies for the second copy, whose generators are called $\widetilde\lambda$. Taking $\delta \sigma=\delta H_\pm^{(0)}=\delta Y^\pm=0$, the gauge parameters close respectively the Lie algebras $[\lambda_1,\lambda_2]_M = \lambda_{12}$ and $[\widetilde{\lambda}_1,\widetilde{\lambda}_2]_M = \widetilde{\lambda}_{12}$, where $\lambda_{12}$ and $\widetilde{\lambda}_{12}$ depend on
\begin{equation}
    Y^\pm_{12} = {Y^\pm_2} \partial_\pm {Y^\pm_1} - {Y^\pm_1} \partial_\pm {Y^\pm_2} \, , \qquad \sigma_{12} = 0 \, , \qquad H_{\pm \, 12}^{(0)} = 0 \, .
\end{equation}
The residual symmetry algebra is given by the direct sum of two Witt algebras generated by $Y^i$, an abelian sector generated by $\sigma$ and another one generated by the boundary vector $H^{(0)}_i$.

If we introduce the following Fourier mode expansions of the symmetry parameters
\begin{equation} \label{sym parameters mode expansions}
    Y^\pm \sim \text{e}^{\text{i} n x^\pm} \, , \quad \sigma \sim \text{e}^{\text{i} p x^+} \text{e}^{\text{i} q x^-} \, , \quad {H_{\pm}^{(0)}} \sim \text{e}^{\text{i} p x^+} \text{e}^{\text{i} q x^-} \, ,
\end{equation}
with $n,p,q \in \mathbb{Z}$, we gather
\begin{equation} \label{asymptotic sym algebra}
    \begin{aligned}
    &[\lambda_n^{Y^\pm},\lambda_m^{Y^\pm}]_M = \text{i} (n-m) \lambda_{n+m}^{Y^\pm} \, , \quad &&[\lambda_n^{Y^\pm},\lambda_m^{Y^\mp}]_M = [\lambda_n^{Y^\pm},\lambda_{pq}^{\sigma}]_M = [\lambda_n^{Y^\pm},\lambda_{pq}^{H_\pm^{(0)}}]_M = 0 \, ,\\
    &[\lambda_{pq}^{\sigma},\lambda_{rs}^{\sigma}]_M = 0 \, , \quad &&[\lambda_{pq}^{\sigma},\lambda_{rs}^{H_\pm^{(0)}}]_M = 0 \, ,\\
    &[\lambda_{pq}^{H_\pm^{(0)}},\lambda_{rs}^{H_\pm^{(0)}}]_M = 0 \, , \quad &&[\lambda_{pq}^{H_\pm^{(0)}},\lambda_{rs}^{H_\mp^{(0)}}]_M = 0 \, .
    \end{aligned}
\end{equation}
In these commutation relations, for example, we have denoted by $\lambda_n^{Y^\pm}$ the gauge parameters where we turned on only $Y^\pm$ expanded as in \eqref{sym parameters mode expansions} and set the other  parameters $\sigma$ and $H^{(0)}_{\pm}$ to zero.

\paragraph{Surface charges}

Using \eqref{gauge symmetry of CS field}, the variations of the boundary data under the above  transformations read
\begin{align} \label{ChernSimons - AdS Variation of Frame Function}
\delta_{(\lambda,\widetilde{\lambda})} \ell_\pm &= Y^\pm \partial_\pm \ell_\pm + 2 \ell_\pm \partial_\pm Y^\pm -\frac{1}{2} \partial_\pm^3 Y^\pm \, , \qquad \delta_{(\lambda,\widetilde{\lambda})} \phi = \sigma \, , \qquad \delta_{(\lambda,\widetilde{\lambda})} k_\pm^{(2n)} = h_\pm^{(2n)} \, ,
\end{align}
with $n \in \mathbb{N}$. In particular, we have that
\begin{equation}
    \delta_{(\lambda,\widetilde{\lambda})} K_\pm^{(0)} = H_\pm^{(0)} \, .
\end{equation}
We summarize the covariant phase space formalism applied to Chern--Simons theories in appendix \ref{app sec: CPSF}. Following \cite{Banados:1994tn}, the surface charges evaluated at fixed value of the time coordinate $t$ are given by
\begin{equation}
\delta Q_\lambda = - \frac{\kappa}{2\pi} \int_0^{2\pi} \text{d}\theta \, \text{tr} \Big( \lambda \, \delta A_\theta \Big) \, , \qquad \delta \widetilde{Q}_{\widetilde{\lambda}} = - \frac{\kappa}{2\pi} \int_0^{2\pi} \text{d}\theta \, \text{tr} \Big( \widetilde{\lambda} \, \delta \widetilde{A}_\theta \Big) \, ,
\end{equation}
where $\kappa = \nicefrac{1}{4\mathcal{G}}$. We provide in \eqref{generic CS charge} a derivation of this result. In three-dimensional WFG gauge, the symplectic structure does not diverge and the charges are thus finite in the limit $\rho \to 0$. Substituting the gauge parameters, one obtains, up to integration by parts,
\begin{equation}
	\delta Q_{\Lambda} = \lim_{\rho \to 0} \left( \delta Q_{\lambda} - \delta\widetilde{Q}_{\widetilde{\lambda}} \right)= - \frac{\kappa}{2\pi} \int_0^{2\pi} \text{d}\theta \Big[ \delta \ell_+ Y^+ - \delta \ell_- Y^- - \delta\phi H^{(0)}_t  +\sigma \delta K^{(0)}_t \Big] \, ,
\end{equation}
where we denoted $ {\Lambda} = (\lambda,\widetilde{\lambda})$. Note that we have used $A_\pm = \tfrac{1}{2} (\pm A_t + A_\theta)$ in this computation. The above variation is manifestly integrable for $\delta \sigma=\delta H_\pm^{(0)}=\delta Y^\pm=0$, therefore the total surface charges are
\begin{equation} \label{total corner charge}\boxed{
   Q_\Lambda = - \frac{\kappa}{2\pi} \int_0^{2\pi} \text{d}\theta \Big[ \ell_+ Y^+ - \ell_- Y^- - \phi H_t^{(0)} + \sigma K_t^{(0)} \Big]\, .}
\end{equation}
The charge associated to $Y$ is conserved. This is not the case for the charges associated to $\sigma$ and $H_t^{(0)}$, as expected as we did not impose any conditions on the conformal factor of the boundary metric and the Weyl connection.

We would like to offer some remarks. Firstly, the fields associated with the higher radial orders of the Weyl structure $k_i^{(2p)}$, for $p \in \mathbb{N}_0$, and the component $K^{(0)}_\phi$ do not contribute to the asymptotic charges. This means that they are pure gauge and so we can set them to zero by a trivial diffeomorphism. Hence, we are allowed to reduce the asymptotic expansion of the Weyl connection to the leading order only, which in turns makes the metric expansion finite. In this way, we recover the gauge \eqref{WFG CS radial expansion of ki} where one can perform a radial gauge fixing as in \eqref{first CS gauge choice}. On the other hand, using \eqref{diffeomorphism required to go the usual FG gauge} to further fix the gauge from WFG to FG would set to zero a physical charge. This does not contradict the FG theorem, which states that the FG gauge is always reachable, and that could happen at the expenses of setting some charges to zero. This is one of the main results of this paper, that proves how the complete FG gauge fixing is exploiting a charged diffeomorphism, and thus it is restricting the set of physical states. Notice therefore that, in the holographic setup where this charge is non-vanishing, the associated current is physical. It was important to perform the asymptotic symmetry analysis to reply to this question, raised in \cite{Ciambelli:2019bzz, Jia:2021hgy}. 

Secondly, in the conformal parametrization \eqref{conformal gauge condition bdy metric}, the Ricci scalar associated to the boundary metric reads as
\begin{equation} \label{bdy scalar curvature}
    R^{(0)} =- 8 \text{e}^{-2 \phi} \partial_+ \partial_- \phi \, ,
\end{equation}
while the Weyl curvature is given by
\begin{equation} \label{bdy Weyl curvature}
    f^{(0)} = \frac{1}{2} \mathcal{E}^{ij} f_{ij}^{(0)} = 2 \text{e}^{-2\phi} \left( \partial_- k_+^{(0)} - \partial_+ k_-^{(0)} \right) .
\end{equation}
In the last expression, we have used \eqref{new Weyl curvature} and introduced $\mathcal{E}_{ij} = \sqrt{-h^{(0)}} \varepsilon_{ij}$ with $\varepsilon_{01} = +1$ and $\mathcal{E}^{ik} \mathcal{E}_{kj} = \delta^i_j$. Then, imposing the additional constraints
\begin{equation}
    k_\pm^{(0)}(x^+,x^-) = \partial_\pm \zeta(x^+,x^-) \, , \qquad \phi(x^+,x^-) \to \phi(x^+,x^-) + \zeta(x^+,x^-) \, ,
\end{equation}
we set $f^{(0)}=0$, and we match the CS boundary conditions \eqref{boundary conditions on A - CS WFG zero order conformal gauge} studied in \cite{Campoleoni:2022wmf}. Nonetheless, in the latter a different framework has been considered,  where the main focus was to obtain a smooth flat limit of the bulk metric, attainable by focusing on a relaxed Bondi gauge. The main difference is that in WFG  the boundary Weyl scalar curvature contributes to the charges,\footnote{The $2$-dimensional Weyl scalar curvature differs from the Levi-Civita scalar curvature only by total derivatives \cite{Ciambelli:2019bzz}.} while in the relaxed Bondi gauge it is the boundary Weyl curvature that is charged, instead. We leave a better understanding of the link between the different gauge relaxations in three-dimensions for future work. 

Finally, we notice that  $\phi$ and $K^{(0)}_t$ are Heisenberg partners in the sense that $\delta_\xi\phi=\sigma$ and $\delta_\xi K^{(0)}_t=H_t^{(0)}$. This implies that they purely come from a corner term in the symplectic potential, as they can  be written as $\D C$, see \eqref{ambiguities def}. In section \ref{sec:HolographicWI} we discuss different variational principles switching on or off these charges.  This is the manifestation that $\phi$ and $K^{(0)}_t$ are kinematic charges, unlike the charges associated to $\ell^\pm$ that are constrained due to Einstein equations.

\paragraph{Charge algebra}
As a last step in this analysis, we show that, under the Poisson bracket, the charges form a projective representation of the asymptotic symmetry algebra \eqref{asymptotic sym algebra}
\begin{equation} \label{charge algebra}
    \{ {Q}_{ {\Lambda}_1}, {Q}_{ {\Lambda}_2}\} = \delta_{ {\Lambda}_2}  {Q}_{ {\Lambda}_1} =  {Q}_{[ {\Lambda}_1, {\Lambda}_2]_M} + \mathcal{K}[{ {\Lambda}_1, {\Lambda}_2}]\, ,
\end{equation}
where $\mathcal{K}[{ {\Lambda}_1, {\Lambda}_2}]$ is the central extension satisfying the $2-$cocycle condition
\begin{equation}
    \mathcal{K}[[{ {\Lambda}_1, {\Lambda}_2}]_M,\Lambda_3] + \mathcal{K}[[{ {\Lambda}_2, {\Lambda}_3}]_M,\Lambda_1] + \mathcal{K}[[{ {\Lambda}_3, {\Lambda}_1}]_M,\Lambda_2] = 0 \, .
\end{equation}
Indeed, we obtain that the $\text{Witt}\oplus\overline{\text{Witt}}$ part gives rise to a double copy Virasoro algebra with
\begin{equation}
    \mathcal{K}[{\Lambda_1^{Y^\pm},\Lambda_2^{Y^\pm}}] = - \frac{\kappa}{4\pi} \int_0^{2\pi} \text{d}\theta \left( Y_1^+ \partial_+^3 Y_2^+ + Y_1^- \partial_-^3 Y_2^- \right) ,
\end{equation}
while the other parts are promoted to affine algebras with central extensions
\begin{equation}
    \mathcal{K}[{\Lambda_1^{\sigma},\Lambda_2^{H_\pm^{(0)}}}] = - \frac{\kappa}{2\pi} \int_0^{2\pi} \text{d}\theta \, \sigma_1 H_{\pm \, 2}^{(0)} \, .
\end{equation}
In terms of the mode expansions \eqref{sym parameters mode expansions}, the charge algebra \eqref{charge algebra} can be written
\begin{equation} \label{charge algebra modes}
    \{ Q_{\Lambda_n^{Y^\pm}} , Q_{\Lambda_m^{Y^\pm}} \} = \text{i} (n-m) Q_{\Lambda_{n+m}^{Y^\pm}} - \text{i} m^3 \frac{c}{12} \delta_{n+m,0} \, , \qquad \{ Q_{\Lambda_{pq}^{\sigma}} , Q_{\Lambda_{rs}}^{H_\pm^{(0)}} \} = \frac{c}{3} \text{e}^{2 \text{i} (q+s) t} \delta_{p+r,q+s} \, ,
\end{equation}
where $c = \nicefrac{3}{2 \mathcal{G}}$ is the Brown--Henneaux central charge \cite{Brown:1986nw}. Note that we find the same time-dependency in the central charge as in \cite{Alessio:2020ioh}. The charge algebra of the phase space is then a double copy of the Virasoro algebra in direct sum with an Heisenberg algebra. This is precisely the same algebra  appearing in the Bondi-Weyl gauge \cite{Geiller:2021vpg} and in the analysis of generic hypersurfaces \cite{Adami:2020ugu,Adami:2022ktn,Adami:2023fbm}.

\section{Holography and variational principle}
\label{sec:HolographicWI}

In this section, we discuss the holographic renormalization of the action principle in second order formalism. We propose different prescriptions of finite counterterms yielding different expressions for the holographic stress tensor and the holographic Weyl current, as well as for the charges. In particular, we show that the above discussion in first order formalism corresponds to a specific choice of counterterms in the action. 

\subsection{Holographic renormalization}
\label{sec:Holographically renormalized action}

\paragraph{Variational principle} In previous works, the holographic renormalization in the WFG gauge has been performed using dimensional regularization \cite{Ciambelli:2019bzz,Jia:2021hgy}. In this section, we revisit this problem using the cut-off regularization. Starting from the WFG solution space \eqref{WFG ansatz of the bulk metric}, the renormalized action is 
\begin{equation}
\begin{split}
    S_{ren}=\frac1{16\pi \mathcal{G}} \int \text{d}^3x \, &(R+2) +\frac{1}{8\pi \mathcal{G}}\int \text{d}^2x \sqrt{-\gamma} \, (K-1) \\
    &+ \frac1{16 \pi \mathcal{G}}\int \text{d}^2x \sqrt{-\gamma} \,k_i \, \gamma^{ij} \, k_j 
    + \frac{\rho^2\log \rho}{16 \pi \mathcal{G}}\int \text{d}^2 x \sqrt{-\gamma} \, \hat{R}^{(0)} \, ,
    \label{renormalized action}
\end{split}
\end{equation}
where we denote $n_\mu$ the normal to constant-$\rho$ hypersurfaces $n_\mu=-\sqrt{-\gamma}\delta_\mu^\rho$, $\gamma_{\mu\nu}=g_{\mu\nu}-n_\mu n_\nu$ the induced metric, and $K = g^{\mu\nu} \nabla_{\mu} n_{\nu}$ the extrinsic curvature. The first term in the right-hand side of \eqref{renormalized action} is the Einstein--Hilbert bulk action, the second is the Gibbons--Hawking--York boundary term, and the last two terms are boundary counterterms ensuring the finiteness of the on-shell action in WFG gauge. The variation of the on-shell action yields 
\begin{equation}
    \delta S_{ren}\approx\int \text{d}^2x \, \Theta_{ren},
    \label{first variation of tha action}
\end{equation}
with, following the prescription \cite{Compere:2008us,Papadimitriou:2005ii,Freidel:2020xyx, Compere:2020lrt} (see also appendix \ref{app sec: CPSF}),  the renormalized symplectic potential\footnote{In an abuse of vocabulary, we refer to $\Theta$ as the symplectic potential, although at this stage it should be called presymplectic, since it is degenerate.}
\begin{equation}
   \Theta_{ren} =\lim_{\rho\to 0} \left[\Theta^\rho_{EH} +\delta\left(\frac{1}{8\pi \mathcal{G}}\sqrt{-\gamma}(K+1)+ \frac1{16 \pi \mathcal{G}}\sqrt{-\gamma}\,k_i \, \gamma^{ij} \, k_j \right)-\partial_i\vartheta _{GHY}^i \right] .
 \end{equation}
 Here, $\Theta^\mu_{EH} = \frac{\sqrt{-g}}{16\pi \mathcal{G}} [ \nabla_\nu (\delta g)^{\mu\nu} - \nabla^\mu {(\delta g)^\nu}_\nu]$ is the Einstein--Hilbert symplectic potential, and $\lim_{\rho\to 0}\vartheta^i _{GHY}=\frac1{16\pi \mathcal{G}} \sqrt{-h^{(0)}} \delta (h_{(0)}^{ij}k^{(0)}_j)$ is the Gibbons--Hawking--York symplectic potential. Explicitly, we have
\begin{align}
     \Theta_{ren} =- \sqrt{-h^{(0)}} \left(\frac{1}{2}T^{ij} \delta h_{ij}^{(0)} - J^i \delta k_i^{(0)} \right) \label{explicit thetaRean}
\end{align}
where the holographic stress tensor $T^{ij}$ and the holographic Weyl current $J^i$ read as\footnote{We denote by the round brackets the symmetrization on the corresponding indices, $A_{(ij)}=\frac12 (A_{ij}+A_{ji})$.}
\begin{align} \label{eq:stresstensor}
    T^{ij} &= -\frac{2}{\sqrt{-h^{(0)}}} \frac{\delta S_{ren}}{\delta h_{ij}^{(0)}} \approx \frac1{8\pi \mathcal{G}}\left( h_{(2)}^{ij}+\frac12 h_{(0)}^{ij} R^{(0)} + \hat \nabla^{(i}_{(0)} k^{j)}_{(0)}\right) ,\\
    J^i &= \frac{1}{\sqrt{-h^{(0)}}} \frac{\delta S_{ren}}{\delta k_i^{(0)}} \approx \frac1{8\pi \mathcal{G}} \, k_{(0)}^i \, .\label{j}
\end{align}
Notice the presence of the $\hat \nabla^{(i}_{{(0)}} k^{j)}_{(0)}$  term in the holographic stress tensor compared to the usual Brown--York expression in FG gauge. In the holographic dictionary, the renormalized symplectic potential \eqref{explicit thetaRean} has the standard form VEV $\times$ $\delta$(sources). The sources correspond to the boundary geometry which, in WFG gauge, is provided by the conformal class of boundary metric $h^{(0)}_{ij}$ and the Weyl connection $k^{(0)}_i$. As a result, in addition to the VEV of the holographic stress tensor associated with the source $h^{(0)}_{ij}$, there is a VEV for the Weyl current operator $J^i$ associated with the source $k^i_{(0)}$. The unsual feature, which appears in three bulk dimensions, is that the Weyl current is given on shell by the Weyl connection \eqref{j}.

\paragraph{Holographic Ward identities} As discussed in section \ref{subsec:WFG charges first order}, the residual gauge diffeomorphisms of the WFG gauge are given by
$\xi^\rho = \rho \omega + \mathcal{O}(\rho^3)$ and $\xi^i = Y^i + \rho^2 \zeta^i + \mathcal{O}(\rho^4)$ at leading order. Under these residual diffeomorphisms, the boundary geometry transforms as
\begin{equation}
 \delta_{(Y, \omega, \zeta)}h_{ij}^{(0)}=\mathcal{L}_{Y} h_{ij}^{(0)}-2 \, \omega \, h_{ij}^{(0)} \,, \qquad \delta_{(Y, \omega, \zeta)} k^{(0)}_i =   \mathcal{L}_{Y} k^{(0)}_i - \partial_i \omega - 2 \, \zeta_i \, ,
 \label{variation sources}
\end{equation}
while the holographic stress tensor and Weyl current transform as\footnote{The currents $J^i$ and $T^{i}{}_{j}$ are by construction of weight $+2$, the dimension of the boundary. Moreover, one has $\nabla_i^{(0)} J^i=\hat \nabla_i^{(0)} J^i$.}
\begin{align} \label{variations vev}
\begin{split}
    \delta_{(Y, \omega, \zeta)} T^{ij} &= \mathcal{L}_{Y} T^{ij} + 4 \, \omega \, T^{ij} -\frac1{8\pi \mathcal{G}}\hat\nabla^{(i}_{(0)} \partial^{j)}\omega +\frac1{8\pi \mathcal{G}}h^{ij}_{(0)}\hat\nabla^{k}_{(0)} \partial_k\omega -\frac1{2\pi \mathcal{G}} k_{(0)}^{(i}\zeta^{i)}\\
    &\quad +\frac1{4\pi \mathcal{G}}h^{ij}_{(0)} k^{(0)}_k\zeta^k \, ,
\end{split}\\
\delta_{(Y, \omega, \zeta)} J^i & =  \mathcal{L}_{Y} J^i+2 \, \omega \, J^i -\frac1{8\pi \mathcal{G}} \partial^i \omega -\frac1{4\pi \mathcal{G}} \zeta^i \, .
\end{align}
The holographic Ward identities are obtained by evaluating the variation of the on-shell action \eqref{first variation of tha action} on the above symmetries. Under the action of boundary diffeomorphisms, we have
\begin{equation}
\begin{split}
    \delta_Y S_{ren} &= - \int \text{d}^2x  \sqrt{-h^{(0)}} \left(\frac{1}{2}T^{ij} \delta_Y h_{ij}^{(0)} - J^i \delta_Y k_i^{(0)} \right) \\
    &= - \int \text{d}^2x  \sqrt{-h^{(0)}} \, Y^j \left(- \nabla_i^{(0)}  {T^i}_j +  J^i f_{ij}^{(0)} + \nabla_i^{(0)} J^i k^{(0)}_j  \right)
\end{split}
\end{equation}
where $f_{ij}^{(0)} = \nabla_i^{(0)} k^{(0)}_j - \nabla_j^{(0)} k^{(0)}_i$ is the Weyl curvature corresponding to the leading order of \eqref{new Weyl curvature}. To obtain the second equality, we used the explicit transformation of the sources in \eqref{variation sources} under boundary diffeomorphisms and integrated by parts to isolate the symmetry parameter $Y^j$. Einstein equations imply that this expression vanishes
\begin{equation}
    \nabla_i^{(0)} {T^i}_j =  J^i \, f_{ij}^{(0)} +  \nabla_i^{(0)} J^i \, k^{(0)}_j, 
    \label{WI for diffeos}
\end{equation}
which yields the invariance of the renormalized on-shell action under boundary diffeomorphisms. Equation \eqref{WI for diffeos} can then be interpreted as the holographic Ward identity associated with boundary diffeomorphisms. Notice the presence of the additional terms in the right-hand side due to the Weyl connection compared to the usual covariant conservation of the stress tensor. 

The variation of the action under boundary Weyl transformations is
\begin{equation}
\begin{split}
    \delta_\omega S_{ren} &=  - \int \text{d}^2x  \sqrt{-h^{(0)}} \left(\frac{1}{2}T^{ij} \delta_\omega h_{ij}^{(0)} - J^i \delta_\omega k_i^{(0)} \right) \\
    &= \int \text{d}^2x  \sqrt{-h^{(0)}} \, \omega  \left(  {T^i}_i + \hat{\nabla}_i^{(0)} J^i \right) \\
    &= \frac{c}{24\pi} \int \text{d}^2x  \sqrt{-h^{(0)}} \, \omega \, \hat{R}^{(0)} \, .
\end{split} \label{holo weyl anomaly}
\end{equation}
In the last equality, we used Einstein equations to make the Weyl--Ricci scalar appear. This shows that the renormalized on-shell action is not invariant under Weyl, which unveils the presence of a holographic Weyl anomaly \cite{Henningson:1998gx,deHaro:2000vlm}. Thus, the anomalous Ward identities associated with Weyl rescalings give
\begin{equation}
     {T^i}_i + \hat{\nabla}_i^{(0)} J^i = \frac{c}{24\pi} \hat{R}^{(0)} \, .
     \label{Weyl WI}
\end{equation} 
Plugging \eqref{Weyl WI} into \eqref{WI for diffeos}, we find a suggestive rewriting of the Ward identity for boundary diffeomorphisms
\begin{equation}
    \hat{\nabla}_i^{(0)} {T^i}_j = J^i f_{ij}+\frac{c}{24\pi} \hat R^{(0)} \, k^{(0)}_j \, .
\end{equation}
As we shall now discuss, the symmetries $\zeta^i$ are pure gauge for the phase space obtained from the holographic renormalization of the action \eqref{renormalized action}, and therefore do not lead to additional holographic Ward identities.

\paragraph{Charges} 
We compute the charges using the covariant phase space formalism \cite{Crnkovic:1986ex,Lee:1990nz,Wald:1993nt,Wald:1999wa,Barnich:2001jy} (see appendix \ref{app sec: CPSF} and, e.g., \cite{Riegler:2017fqv,Compere:2018aar,Ruzziconi:2019pzd,Ruzziconi:2020cjt,Ciambelli:2022vot} for recent reviews). 
First, we contract the renormalized symplectic current
\begin{equation}
    \omega_{ren} = \delta \Theta_{ren} = - \frac{1}{2} \delta (\sqrt{-h^{(0)}} T^{ij}) \wedge \delta h_{ij}^{(0)} + \delta (\sqrt{-h^{(0)}}  J^i) \wedge \delta k_i^{(0)}  
\end{equation}
with the symmetries encoded in the variations \eqref{variation sources} and \eqref{variations vev}. Then, to compare with the analysis of section \ref{subsec:WFG charges first order} in Chern--Simons formulation, we choose the conformal gauge \eqref{conformal gauge condition bdy metric} for the boundary metric. After the redefinitions \eqref{K_pm^0} and \eqref{field dep redef}, assuming $\delta \sigma=\delta Y=0$, we find that the canonical charge is finite and given by
\begin{equation}
    Q_\xi = \int q_\xi = \frac1{8\pi \mathcal{G}}\int^{2\pi}_0 \text{d}\theta \left[Y^+\, \ell_+-Y^-\, \ell_- +\partial_t\sigma\,\phi -\sigma \,\partial_t\phi\right] .
    \label{asymptotic charges}
\end{equation} 
Notably, the charges \eqref{asymptotic charges} match precisely with the expressions found in the FG gauge in \cite{Alessio:2020ioh}. In contrast with the analysis of section \ref{subsec:WFG charges first order}, the charge associated with the symmetry $\zeta^i$ vanishes, which implies that this residual gauge diffeomorphism is now pure gauge and can be quotiented out. Note that this result holds for arbitrary boundary metric and not only in the conformal gauge. We observe that the charge algebra is unaffected, and again given by two copies of the Virasoro algebra in direct sum with a Heisenberg algebra. However, here this Heisenberg factor is realized solely by the zero mode of the conformal factor and its first derivative, while in \eqref{total corner charge} the Heisenberg pair comes from different boundary fields.

Therefore, the phase space controlled by the action after holographic renormalization \eqref{renormalized action} differs from the one in the Chern--Simons formulation discussed in section \ref{subsec:WFG charges first order}. In this framework, the Weyl connection can be consistently set to zero. The holographic interpretation is that there are no observables that are sensitive to the latter, and thus the Weyl covariantization of the boundary geometry can be achieved without enlarging the physical phase space. On the other hand, this is not the case in section \ref{sec: Higher order}, where there is a non-vanishing charge, and we would like to understand how to derive this result in second-order formulation, to gain a better holographic perspective.

It is well known that the renormalized action \eqref{renormalized action} is not uniquely defined: one can add finite counterterms. As we shall now explain, a judicious choice of boundary or corner counterterms will allow to recover the phase space found in the Chern--Simons formulation. 

\subsection{Boundary counterterm}
\label{sec:Boundary counter-term}

We now connect the holographic discussion in the WFG gauge presented in section \ref{sec:Holographically renormalized action} with the results of section \ref{subsec:WFG charges first order}. In the holographic renormalization procedure, one can always add a finite boundary counterterm to the action \eqref{renormalized action} as 
\begin{equation}
    \bar{S}_{ren} = S_{ren} + S_\circ, \qquad S_\circ = \int \text{d}^2x \, L_\circ[h^{(0)}_{ij}, k_i^{(0)}]
    \label{tilde renormalized action}
\end{equation}
where $L_\circ[h^{(0)}_{ij}, k_i^{(0)}]$ is a boundary Lagrangian involving the boundary geometry. Defining 
\begin{equation}
T^{ij}_o = -\frac{2}{\sqrt{-h^{(0)}}} \frac{\delta S_o}{\delta h_{ij}^{(0)}}, \qquad J^i_o =  \frac{1}{\sqrt{-h^{(0)}}} \frac{\delta  S_o}{\delta k_i^{(0)}}
\end{equation}
and using the standard procedure to treat boundary terms \cite{Compere:2008us,Papadimitriou:2005ii,Freidel:2020xyx, Compere:2020lrt}, the variation of the renormalized on-shell action reads as
\begin{equation}
    \delta \bar{S}_{ren} \approx \int \text{d}^2x \, \bar{\Theta}_{ren}
\end{equation}
with 
\begin{equation}
\bar{\Theta}_{ren} =- \sqrt{-h^{(0)}} \left(\frac{1}{2}\bar{T}^{ij} \delta h_{ij}^{(0)} - \bar{J}^i \delta k_i^{(0)} \right) 
\label{theta ren 2}.
\end{equation}
The new currents are given by
 $\bar{T}^{ij} = T^{ij} + T^{ij}_\circ$ and $\bar{J}^i = J^i + J^i_\circ$. Choosing the boundary term 
\begin{equation}
    L_\circ = \lim_{\rho \to 0} \left[\frac{1}{16\pi \mathcal{G}} k_i \gamma^{ij} \partial_j \sqrt{-\gamma} \right] = \frac{1}{16\pi \mathcal{G}} k_i^{(0)} h^{ij}_{(0)} \partial_j \sqrt{-h^{(0)}} \, ,
    \label{boundary langrangian prime}
\end{equation}
we find explicitly 
\begin{align}
  \bar  T^{ij}&=T^{ij} +J^{(i}\partial^{j)}\sqrt{-h^{(0)}}+\frac12 h^{ij}\nabla_kJ^k\,, \qquad     \bar J^i=J^i+\frac1{16\pi \mathcal{G}}\partial^i\log\sqrt{-h^{(0)}} \, .
  \label{barT}
\end{align}
Notice that the boundary Lagrangian \eqref{boundary langrangian prime} is not covariant with respect to boundary diffeomorphisms. However, the phase space associated with this choice allows one to reproduce the results of the Chern--Simons formulation in section \ref{subsec:WFG charges first order}. Indeed, starting from \eqref{theta ren 2}
, one reproduces exactly the charge expressions \eqref{total corner charge} after imposing the conformal gauge condition \eqref{conformal gauge condition bdy metric} and performing the redefinitions \eqref{K_pm^0} and \eqref{field dep redef}. In particular, with this choice of counterterms one finds new symmetries compared to the FG gauge, as explained in the CS section \ref{sec: Higher order}. 

We see therefore that there is a trade-off: one can promote the Weyl connection to a physical source at the expense of introducing a non-covariant boundary Lagrangian. In this framework, we are reproducing the holographic aspects explored in \cite{Ciambelli:2019bzz, Jia:2021hgy}, where it was proposed that the Weyl connection should be treated as an independent source. In this setup, the extra current should be added to the boundary partition function, and thus there will be non-trivial correlation functions involving it. This is an open road to explore further.

\subsection{Corner counterterm}

Interestingly, there is another procedure to extract the symplectic potential from the variational principle \cite{McNees:2023tus}, which yields the charges of section \ref{subsec:WFG charges first order}. This procedure amounts to include corner contributions of the bulk symplectic potential.
 Concretely the renormalized symplectic potential is 
\begin{equation}
  \tilde  \Theta_{ren} = \lim_{\rho\to 0}\left[\Theta^\rho_{EH} +\partial_i \int \text{d}\rho \,\Theta^i_{EH} + \delta  L_b \right]
\end{equation}
where $L_b$ are the boundary Lagrangians (that can include corner Lagrangians). This differs from the prescriptions applied in sections \ref{sec:Holographically renormalized action} and \ref{sec:Boundary counter-term} for which $C$ defined in eq. \eqref{ambiguities def} is equal to minus the symplectic potentials associated to the boundary Lagrangians. 

Let us describe the choice of boundary and corner Lagrangians for the case of interest. 
We add a finite corner term to the renormalized action \eqref{renormalized action}:
\begin{equation}
    \tilde S_{ren} = S_{ren} + S_C, \qquad S_C = \int \text{d}^2x \, \partial_i L_C^i [h^{(0)}_{ij}, k_i^{(0)}]
    \label{corner action}
\end{equation}
where $L_C^i [h^{(0)}_{ij}, k_i^{(0)}]$ is a corner Lagrangian involving the boundary geometry. We take
\begin{equation}
    L_C^i = \lim_{\rho \to 0} \left[- \frac{1}{16\pi \mathcal{G}} \sqrt{-\gamma} \gamma^{ij} k_j \right] = - \frac{1}{16\pi \mathcal{G}} \sqrt{-h^{(0)}} h^{ij}_{(0)} k_j^{(0)}. \label{corner lagrangian}
\end{equation}
%
This corner Lagrangian \eqref{corner lagrangian} has the advantage of being covariant with respect to boundary diffeomorphisms. The corner contribution of the bulk symplectic potential $\Theta_{EH}$ turns out to be $\delta$-exact, $\lim_{\rho\to 0}\partial_i\Theta_{EH}^i=\delta\left(-\frac1{16 \pi \mathcal G} \log \rho \sqrt{-h^{(0)}} \hat R^{(0)}\right)$, precisely canceling the log-term contribution from the boundary Lagrangians.
Then the renormalized symplectic term is
\begin{equation}
    \begin{split}
    \tilde \Theta_{ren} &= \lim_{\rho \to 0}\bigg[\Theta^\rho_{EH} +\partial_i \int \text{d}\rho \, \Theta^i_{EH} +\frac{1}{8\pi \mathcal G}\delta\bigg(\sqrt{-\gamma} \, (K+1)+ \frac1{2}\sqrt{-\gamma} \, k_i \gamma^{ij} k_j\\
    & \quad + \frac{\rho^2\log \rho}{2}\int \text{d}^2 x \sqrt{-\gamma} \, \hat{R}^{(0)} -\frac12\partial_i (\sqrt{-\gamma} \, \gamma^{ij} k_i)\bigg) \bigg]\\
    &= \sqrt{-h^{(0)}} \left( -\frac12 \tilde T^{ij}\delta h_{ij}^{(0)} +J^i\delta K_i^{(0)}\right) , \label{tilde ren}
    \end{split}
\end{equation}
where 
\begin{align}
  \tilde  T^{ij}&=T^{ij} +\frac12 h^{ij}_{(0)} \hat \nabla_k^{(0)} J^k\,, \qquad     K_i^{(0)} = k^{(0)}_i-\frac12 \, \partial_i\ln\sqrt{-h^{(0)}} \,.
\end{align} 
These transform under the residual symmetries $\xi^\rho = \rho \omega + \mathcal{O}(\rho^3)$ and $\xi^i = Y^i + \rho^2 \zeta^i + \mathcal{O}(\rho^4)$ as
\begin{align}
    \delta_{(Y, \omega, \zeta)} K^{(0)}_i &=  \mathcal{L}_{Y} K^{(0)}_i - 2 \, \zeta_i-\frac12 \partial_i \partial_j Y^j\,,\\
    \begin{split}
    \delta_{(Y, \omega, \zeta)}\tilde  T^{ij} &= \mathcal{L}_{Y} \tilde T^{ij}+ 4 \, \omega \, \tilde T^{ij} - \frac1{8\pi \mathcal{G}}\bigg( \hat\nabla^{(i}_{(0)} \partial^{j)}\omega -\frac12 h^{ij}_{(0)}\hat\nabla^{k}_{(0)} \partial_k\omega+ 4k_{(0)}^{(i}\zeta^{i)}\\
    &\quad + h_{(0)}^{ij} \left( \hat\nabla_{k}^{(0)}\zeta^k -2 k_k^{(0)}\zeta^k \right) \bigg) \, .
    %
    \end{split}
\end{align}

With this prescription, the holographic interpretation of \eqref{tilde ren} as VEV $\times$ $\delta$(sources) leads to the identification of $h_{ij}^{(0)}$ and $K_i^{(0)}$ (called Weyl source from now on) as sources. However, the latter is not a Weyl connection. Rather, it has been dressed by a Weyl pure-gauge shift, so that it is now Weyl invariant. It would be interesting to study the holographic repercussions of this choice, where the two sources transform independently of each other, similar to what happens when adding a $U(1)$ gauge field in the bulk -- and indeed one could perhaps gain insights by treating Einstein--Maxwell bulks with this analogy. In particular, the transformation of the Weyl source is now reminiscent of a one-form symmetry, and thus, from the dual perspective, there could be Wilson line associated to it. Understanding this from the intrinsic boundary field-theoretically perspective is a thrilling open question.

Repeating the same manipulations as above in conformal gauge, the phase space associated with the symplectic potential \eqref{tilde ren} reproduces exactly the result \eqref{total corner charge} of the charge in the first order formulation. Finally, since the boundary terms in the action and the symplectic corner terms only contribute to $\delta$-exact terms in the symplectic potential \eqref{tilde ren},
the symplectic current coincides precisely with the Einstein--Hilbert symplectic current. Therefore, the charges obtained here and in the first order formulation coincide with those obtained in the covariant phase space formalism, without renormalization or any modification related to boundary terms in the action.

\section{Conclusions} \label{sec: Conclusions}
In this work, we have argued that the Weyl--Fefferman--Graham gauge is more suitable than the standard Fefferman--Graham gauge for two reasons: $(i)$ the Weyl rescalings at the boundary are induced by purely radial bulk diffeomorphisms \eqref{Weyl in WFG}. This contrasts with the Weyl rescalings in Fefferman--Graham gauge which necessarily involve a mixing with the transverse components. As a consequence, each order in the radial expansion transforms with a well-defined Weyl weight, see eq. \eqref{WeylcovD}. $(ii)$ The Weyl--Fefferman--Graham
gauge induces the full Weyl geometry (i.e. a conformal class of metrics together with a Weyl connection) at the boundary. This allows us to exploit the full Weyl covariance of the boundary theory. In particular, the Weyl anomaly derived in \eqref{holo weyl anomaly} is Weyl covariant since it involves the Weyl--Ricci scalar. 

To obtain the WFG gauge, one has to relax the FG gauge and allow fluctuations of $k_i(\rho, x)$ in \eqref{WFG ansatz of the bulk metric} on the phase space. Relaxing this gauge condition implies the appearance of a larger set of independent residual gauge symmetries in the subleading order of the transverse bulk diffeomorphisms, $\zeta^\pm$ in \eqref{AKV CS WFG}. The novelty of our work is to compute the charges associated with the residual diffeomorphisms of the WFG gauge. We have shown that, for a suitable choice of symplectic structure, these additional residual symmetries are non vanishing and therefore carry physical information. These results were discussed both in first order Chern--Simons formulation of gravity, as well as in second order metric formulation. The new charge associated to the Weyl connection combines with the Weyl charge to form a Heisenberg algebra. These two charges are then purely corner charges unconstrained by the equations of motion. The derivation of the symplectic structure from the  renormalized action has also been provided. These results confirm previous works \cite{Grumiller:2016pqb,Grumiller:2017sjh,Grumiller:2019fmp,Adami:2020ugu,Ciambelli:2020ftk,Ciambelli:2020eba,Ruzziconi:2020wrb,Geiller:2021vpg,Campoleoni:2022wmf,Adami:2023fbm} advocating that, in presence of boundaries, the complete gauge-fixing might eliminate potentially interesting physical degrees of freedom. 

Investigating the implications of these additional physical symmetries for the dual field theory is an interesting question for future endeavour. A first glance revealed that the holographic interpretation turned out to be polyvalent: either the Weyl connection is a new source, but there is a non-covariant boundary Lagrangian, or the new source is a Weyl-invariant combination of the Weyl connection and the metric, and the boundary Lagrangian employed is covariant. We have commented both instances, and we in particular noted that the Weyl connection and the Weyl-source transformation laws under $H^{(0)}$ are reminiscent of a one-form symmetry. When the associated charge is non-vanishing, then there are physical states at the boundary sensitive to this operator, which can be interpreted as a Weyl Wilson line, and would be a primer in the realization of higher form symmetries from bottom-up holography. We refer e.g. to \cite{Ciambelli:2020ftk,Ciambelli:2020eba,Campoleoni:2022wmf} for a holographic interpretation of symmetries obtained beyond the complete gauge fixing.

While the analysis here has been performed in three dimensions of bulk, we expect similar results to hold for higher dimensions. For instance, it would be worth extending the systematic phase space analysis provided in \cite{Fiorucci:2020xto} for the WFG gauge. While in $3$-dimensional bulks the boundary is always conformally flat and thus the full conformal isometries group is realized, in higher dimensions there are situations in which the boundary is not conformally flat (e.g. $3$-dimensional boundaries with non-vanishing Cotton tensor), and thus the conformal isometries group is smaller. Nonetheless, the dual theory enjoys Weyl covariance, because this symmetry arises solely from the fact that the boundary sits at conformal infinity. This means that we are dealing in these cases with a conformal field theory on curved spaces with Weyl symmetry, i.e., a Weyl field theory. Extending our analysis to higher dimensions could potentially shed light on these theories, of which little is known.

As a general statement, we have shown that new charges might arise from boundary Lagrangians. Therefore, this raises the important question of classifying new charges associated to choices of symplectic spaces. As a general guideline, one could argue that the more physical charges the better, as this would lead to larger algebras that are more powerful to organize observables of the theory. This specific example treated here is therefore opening the door to a more fundamental problem, which is the classification of charges steaming from partial gauge fixings. While this is an interesting avenue in the theory of asymptotic symmetries, we foresee far-reaching repercussions in both AdS and flat holography, yet to be unveiled.

\paragraph{Acknowledgements} We would like to thank Francesco Alessio, Glenn Barnich, Andrea Campoleoni, Rob Leigh, Shahin Sheikh-Jabbari and Weizhen Jia for useful discussions. Research at Perimeter Institute is supported in part by the Government of Canada through the Department of Innovation, Science and Economic Development Canada and by the Province of Ontario through the Ministry of Colleges and Universities. AD is supported by the Fonds de la Recherche Scientifique - FNRS under Grants No.\ FC.41161. RR was partially supported by the Austrian Science Fund (FWF), project P 32581-N, and by the Titchmarsh Research Fellowship in Mathematical Physics at the University of Oxford. RR thanks Perimeter Institute for its hospitality where the last stages of this work were completed.

\appendix

\section{Conventions}

\subsection{Chern--Simons formulation} \label{app sec: Conventions CS}

In this appendix, we recall the salient features of the Chern--Simons (CS) formulation of three-dimensional anti-de Sitter (AdS$_3$) Einstein gravity. This allows us to fix the notations used in main body of the paper. The isometry algebra of AdS$_3$ is the $\mathfrak{so}(2,2)$ algebra
\begin{equation} \label{isometry AdS3}
[M_B,M_C] = \epsilon_{BCD} \, M^D , \quad
[M_B,P_C] = \epsilon_{BCD} \, P^D , \quad
[P_B,P_C] = \left( \frac{\mathcal{G}}{\ell} \right)^2 \epsilon_{BCD} \, M^D \, ,
\end{equation}
where $P_B$ and $M_B$ denote, respectively, the transvection and Lorentz generators. The latter are related to the customary Lorentz generators as
\begin{equation}
    M_B = \frac{1}{2} \, \epsilon_{BCD} \, M^{CD} \, .
\end{equation}
In the above algebra, $\mathcal{G}$ is Newton's constant and $\ell$ denotes the AdS radius while we choose the convention $\epsilon^{012} = 1$ for the Levi-Civita symbol. Labelling by $\mu,\nu,\dots$ the bulk base-manifold indices and introducing the $\mathfrak{so}(2,2)-$valued differential one-form
\begin{equation} \label{so(2,2)-connection}
\mathscr{A} = \left( \frac{1}{\mathcal{G}} {e_\mu}^B P_B + {\omega_\mu}^B M_B \right) \text{d} x^\mu \, ,
\end{equation}
where ${e_\mu}^B$ is the bulk dreibein and ${\omega_\mu}^B$ is its associated dualized spin connection, one can rewrite the $3-$dimensional Einstein--Hilbert action as a CS action \cite{Achucarro:1986uwr,Witten:1988hc}
\begin{equation} \label{CS_generic}
S_{EH} = \frac{1}{16\pi} \int_\mathcal{M} \text{Tr} \left(\mathscr{A} \wedge \text{d}\mathscr{A} + \frac{2}{3} \, \mathscr{A} \wedge \mathscr{A} \wedge \mathscr{A} \right) .
\end{equation}
The operator $\text{d}$ is the exterior derivative on the manifold $\mathcal{M} = \text{AdS}_{3}$, such that $\text{d}^2=0$. In this expression we introduced the Killing metric
\begin{equation}
\text{Tr} \left(M_B M_C\right) = \text{Tr} \left(P_B P_C\right) = 0 \, , \qquad 
\text{Tr} \left(M_B P_C\right) = \eta_{BC} \, ,
\end{equation}
with $\eta_{BC}$ the Minkowski metric whose signature is $(-, +, +)$, and used the Cartan equation
\begin{equation} \label{Cartan equation}
    \text{d}e^B + \epsilon^{BCD} \, \omega_C \wedge e_D = 0 \, .
\end{equation}
For negative cosmological constant, $\Lambda = - \ell^{-2} < 0$ (in the main body of the paper we set $\ell=1$), one can take advantage of the isomorphism $\mathfrak{so}(2,2) \cong \mathfrak{sl}(2,\mathbb{R})\oplus \mathfrak{sl}(2,\mathbb{R})$ to rewrite \eqref{CS_generic} as
\begin{equation} \label{S_EH}
S_{EH} = S_{CS}[A] - S_{CS}[\widetilde{A}] \, ,
\end{equation}
with
\begin{equation} \label{actionCS}
S_{CS}[A] = \frac{\kappa}{4 \pi} \int_\mathcal{M} \text{tr} \left( A \wedge \text{d}A + \frac{2}{3} \, A \wedge A \wedge A \right)
\end{equation}
where $\kappa = \nicefrac{\ell}{4\mathcal{G}}$, and a similar expression holds for $\widetilde{A}$. In the latter we have introduced the $\mathfrak{sl}(2,\mathbb{R})-$valued gauge connections $A = A^B J_B$, where we introduced the $\mathfrak{sl}(2,\mathbb{R})$ generators
\begin{equation}
[{J_B},{J_C}] = {\epsilon_{BC}}^D \, J_D \, , \qquad 
\text{tr} (J_B J_C) = \frac{1}{2} \, \eta_{BC} \, .
\end{equation}
In terms of the $\mathfrak{so}(2,2)$ generators, they can be written as
\begin{equation}
{J_B} = \frac{1}{2} \left(M_B + \frac{\ell}{\mathcal{G}} \, P_B \right) .
\end{equation}
The dreibein and the spin connection are related to the CS forms $A$ and $\widetilde{A}$ via
\begin{equation} \label{A-E}
A^B = \omega^B + \frac{1}{\ell} \, e^B \, , \qquad \widetilde{A}^B = \omega^B - \frac{1}{\ell} \, e^B \, .
\end{equation}
These are the main quantities that we scrupulously analyze in the paper.

\subsection{Covariant phase space formalism} \label{app sec: CPSF}

In section \ref{sec: Higher order}, we determined the solution space, the asymptotic symmetries and the associated corner charges in the CS formulation. To do this, we could have used methods from the Hamiltonian approach as in, e.g., \cite{Regge:1974zd, crnkovic1987covariant, gawedzki1991classical, Banados:1994tn, Coussaert:1995zp, Banados:1998gg, Henneaux:1999ib, Bunster:2014mua,Perez:2014pya,Riegler:2017fqv}. Inspired by the metric formulation, we instead used the covariant phase space formalism. The latter was introduced in \cite{GAWEDZKI1972307,kijowski1973finite,kijowski1976canonical}, refined in \cite{Lee:1990nz,Wald:1993nt,Wald:1999wa,Barnich:2001jy}\footnote{See, e.g., \cite{Compere:2018aar,Ruzziconi:2019pzd,Ciambelli:2022vot} for pedagogical reviews.}, and it focuses on the Lagrangian approach. Its main idea is to put together spacetime and phase space.

The differentiable manifold that we consider is the three-dimensional AdS spacetime, $\mathcal{M} = \text{AdS}_{3}$. In appendix \ref{app sec: Conventions CS}, we have introduced the exterior derivative, $\text{d}$, on this manifold. We denote $\iota$ its interior product. The Lie derivative along the flow of a diffeomorphism $\xi \in T \mathcal{M}$ is
\begin{equation} \label{Lie derivative diffeo}
    \mathcal{L}_\xi = \text{d} \iota_\xi + \iota_\xi \text{d} \, .
\end{equation}
One can put together this calculus with the one on field space $\Gamma$. We denote by $\delta$ and $I$ the exterior derivative and the interior product on the latter, respectively, with $\delta^2=0$. Given a vector field $V \in T \Gamma$, we analogously have
\begin{equation}
    \mathfrak{L}_V = \delta I_V + I_V \delta \, .
\end{equation}

The CS Lagrangian form can be derived from its associated action principle \eqref{actionCS},
\begin{equation}
    S_{CS} = \int_\mathcal{M} L \, , \qquad L = \frac{\kappa}{4\pi} \, \text{tr} \left( A \wedge \text{d}A + \frac{2}{3} \, A \wedge A \wedge A \right) .
\end{equation}
An arbitrary field variation $A \to A + \delta A$ of the latter yields, after iterative applications of the inverse Leibniz rule,
\begin{equation} \label{local presymp pot def}
    \delta L = (\text{eom}) \delta A + \text{d} \Theta \, ,
\end{equation}
where $(\text{eom})$ denotes the equations of motion of the theory
\begin{equation} \label{eom CS}
    (\text{eom}) = \text{d}A + A \wedge A \approx 0 \, ,
\end{equation}
and the symbol $\approx$ indicates that we are on-shell of the equations of motion.
Here, $\Theta$ is the local symplectic potential form
\begin{equation}
    \Theta = - \frac{\kappa}{4\pi} \, \text{tr} \left( A \wedge \delta A \right) \, .
\end{equation}
We define the local symplectic two-form as
\begin{equation} \label{local presymp form def}
    \omega = \delta \Theta = - \frac{\kappa}{4\pi} \, \text{tr} \left( \delta A \wedge \delta A \right) \, .
\end{equation}
This local expression can be integrated on an arbitrary Cauchy slice $\Sigma \subset \mathcal{M}$ to give the symplectic two-form
\begin{equation} \label{presymp form def}
    \Omega = \int_\Sigma \omega \, .
\end{equation}
In particular, we consider that the $3-$manifold $\mathcal{M} = \mathbb{R} \times \Sigma$ has the topology of a cylinder, foliated by the spatial slices, such that the boundary of the $2-$manifold $\Sigma$ is a circle, $\partial \Sigma~=~S^1$.

Specifying a bulk Lagrangian is not enough to completely determine the theory. Indeed, different choices of boundary Lagrangians give different associated charges. In the following, we keep using the language of ambiguities, for historical reasons, with the understanding that different choices of the latter truly imply different physical theories, as indeed our holographic analysis reveals.
The symplectic potential possesses two types of ambiguities in its definition \eqref{local presymp pot def}, which do not affect the equations of motion,
\begin{equation} \label{ambiguities def}
    \Theta \to \Theta + \delta B - \text{d} C \, .
\end{equation}
Adding a boundary term to the bulk Lagrangian $L \to L + \text{d}B$ leads to the first type of ambiguity ($B$) whose contribution to the symplectic form $\omega$ vanishes since $\delta^2=0$. The nilpotent aspect and the definition of $\Theta$ as a boundary term in $\delta L$ yields the second type of ambiguity ($C$). It is worth noting that the latter alters $\omega$,
\begin{equation}
    \omega \to \omega - \delta \text{d} C = \omega + \text{d} \delta C =: \omega + \text{d} \omega_C \, ,
\end{equation}
reflecting our ignorance on how to select the boundary terms $\omega_C$, called corner terms, in the symplectic form. This is related to the corner proposal \cite{Donnelly:2016auv,Speranza:2017gxd,Geiller:2017whh,Geiller:2017xad,Freidel:2020xyx,Freidel:2020svx,Freidel:2020ayo,Donnelly:2020xgu,Ciambelli:2021vnn,Freidel:2021cjp,Ciambelli:2021nmv,Ciambelli:2022cfr,McNees:2023tus}. Ambiguities can be used to renormalize the symplectic potential whenever the latter diverges as one approaches the boundary \cite{Bianchi:2001kw,deHaro:2000vlm,Skenderis:2002wp,Hollands:2005ya,Mann:2005yr,Compere:2008us,Papadimitriou:2010as,Compere:2018ylh, Freidel:2019ohg, Compere:2019bua, Compere:2020lrt, Chandrasekaran:2021vyu, Geiller:2022vto, McNees:2023tus, Campoleoni:2023eqp}, and can be further used to restore integrability \cite{Adami:2021sko,Geiller:2021vpg}. It turns out that such a renormalization procedure is always possible at the level of the symplectic potential as recently shown in \cite{Freidel:2019ohg,Geiller:2022vto,McNees:2023tus}.

In the main text, we are using two different prescriptions to compute the charges. The first one is due to Comp\`ere and Marolf \cite{Compere:2008us}. It takes $B$ to be the boundary Lagrangians that are used for the renormalization of the action and $C$ to be minus the boundary symplectic potential associated to B.  The other prescription \cite{McNees:2023tus} also takes the same $B$ but $C$ to be the corner contributions of the bulk symplectic potential, and hence solely relies on the bulk symplectic potential to compute the charges.

In the presence of gauge symmetries, which is the case we are interested in, Noether's second theorem applies, which associates a codimension-2 conserved quantity to a given local symmetry. The gauge symmetry of the CS field is
\begin{equation}
    \delta_\lambda A = I_{V_\lambda} \delta A = \text{d}\lambda + [A,\lambda] \, .
\end{equation}
Applying Noether's second theorem to this symmetry
\begin{equation}
    I_{V_\lambda} \Omega = - \delta Q_\lambda \, ,
\end{equation}
one can obtain the associated on-shell corner charges
\begin{equation} \label{generic CS charge}
    \delta Q_\lambda \approx \int_{S^1} \delta q_\lambda \, , \qquad \delta q_\lambda = - \frac{\kappa}{2\pi} \, \text{tr} \left( \lambda \, \delta A \right) .
\end{equation}
In three dimensions, we can always find integrable slicings for the charges \cite{Adami:2020ugu, Alessio:2020ioh, Adami:2020amw, Ruzziconi:2020wrb, Geiller:2021vpg, Adami:2021nnf}. This is related to the fact that, in this case, there is no propagation of local degrees of freedom.

\section{Second copy} \label{app sec: Second copy sl(2,R)}

In this appendix, we collect the main results of the second copy of the CS fields, since in section \ref{sec: Higher order} we exclusively focus on the first copy in order to compute the asymptotic symmetries and the charges. The bulk CS connections using our boundary conditions are
\begin{subequations} \label{boundary conditions on tildeA - CS WFG zero order conformal gauge}
\begin{align}
    \widetilde{A}_\rho &= \frac{1}{\rho} L_0 + 2 \sqrt{2} \rho^2 \text{e}^{-\phi} \left( k_-^{(2)} L_1 - k_+^{(2)} L_{-1} \right) + \mathcal{O}(\rho^3) \, ,\\ 
    %
    \widetilde{A}_- &= - \frac{1}{\rho} \sqrt{2} \text{e}^\phi L_{-1} - \left( 2 k_-^{(0)} - \partial_- \phi \right) L_0 - \sqrt{2} \rho \text{e}^{-\phi} h_{--}^{(2)} L_{1} - 2 \rho^2 k_-^{(2)} L_0 + \mathcal{O}(\rho^3) \, ,\\
    \widetilde{A}_+ &= - \partial_+ \phi L_0 + \sqrt{2} \rho \text{e}^{-\phi} \partial_+ K_-^{(0)} L_{1} + \mathcal{O}(\rho^3) \, .
\end{align}
\end{subequations}
If we expand the gauge parameters of the second copy as
\begin{equation}
\widetilde{\lambda}(\rho,x^+,x^-) = \sum^{+1}_{B = -1} \widetilde{\epsilon}^B(\rho,x^+,x^-) L_B \, ,
\end{equation}
one obtains that
\begin{subequations} \label{ChernSimons - tildeA gauge parameters}
    \begin{align}
        \widetilde{\epsilon}^{-1} &= - \frac{\sqrt{2}}{\rho} \text{e}^{\phi} Y^- + \mathcal{O}(\rho^3) \, ,\\
        \widetilde{\epsilon}^{1} &= \frac{\rho}{\sqrt{2}} \text{e}^{-\phi} \Big[ \partial_-^2 Y^- + 2 H_-^{(0)} - 2 \Big( \ell_- Y^- - (K_-^{(0)})^2 Y^- - K_-^{(0)} \partial_- Y^- \Big) \Big] + \mathcal{O}(\rho^3) \, ,\\
        \widetilde{\epsilon}^0 &= - \sigma + \partial_- Y^- - 2 Y^- K_-^{(0)} \, .
    \end{align}
\end{subequations}
%


\bibliographystyle{JHEP.bst}
\providecommand{\href}[2]{#2}\begingroup\raggedright\endgroup

\end{document}